\documentclass[a4paper,twocolumn,superscriptaddress,11pt]{quantumarticle}
\pdfoutput=1
\usepackage[utf8]{inputenc}
\usepackage[english]{babel}
\usepackage[T1]{fontenc}
\usepackage{amsmath}
\usepackage{hyperref}

\usepackage{tikz}
\usepackage{lipsum}

\usepackage{dsfont}
\usepackage[normalem]{ulem}
\usepackage{mathtools}
\usepackage[makeroom]{cancel}
\usepackage{bbm}

\usepackage[sort&compress, numbers]{natbib}
\usepackage{amsfonts}
\usepackage{amsmath}
\usepackage{color}
\usepackage{amssymb}
\usepackage{graphicx}
\definecolor{martin}{rgb}{0,.4,1}

\definecolor{henrik}{rgb}{1,.4,0}

\newcommand{\tr}{\mathrm{Tr}} %old
\newcommand{\Tr}{\mathrm{Tr}} %new

\newcommand{\id}{\mathbbm{1}}

\newcommand{\ket}[1]{\left.\left|{#1}\right.\right\rangle}

\newcommand{\bra}[1]{\left.\left\langle{#1}\right.\right|}

\newcommand{\braket}[2]{\left\langle #1 \middle| #2 \right\rangle}

\newcommand{\average}[1]{\left\langle #1\right\rangle}

\newcommand{\deq}[1]{\stackrel{#1}{=}}
\newcommand{\bigo}[1]{\mathcal{O}\left (#1\right)}

\newcommand{\rom}[1]{\uppercase\expandafter{\romannumeral #1\relax}}
\newcommand{\norbra}[1]{\left( #1\right)}
\newcommand{\sqrbra}[1]{\left[ #1\right]}
\newcommand{\curbra}[1]{\left\{ #1\right\}}
\newcommand{\lind}{\mathcal{L}}
\newcommand{\J}{\mathbb{J}}

\usepackage{amssymb}             % AMS Math

\begin{document}

\title{Thermodynamic length in open quantum systems}
\date{\today}
\author{Matteo Scandi}
\affiliation{Max-Planck-Institut f\"ur Quantenoptik, D-85748 Garching, Germany}

\author{Mart\' i Perarnau-Llobet}
\affiliation{Max-Planck-Institut f\"ur Quantenoptik, D-85748 Garching, Germany}

%\orcid{0000-0003-0290-4698}
%\thanks{You can use the \texttt{\textbackslash{}email}, \texttt{\textbackslash{}homepage}, and \texttt{\textbackslash{}thanks} commands to add additional information for the preceding \texttt{\textbackslash{}author}. If applicable, this can also be used to indicate that a work has previously been published in conference proceedings.}

\maketitle

\begin{abstract}
  The dissipation generated during a quasistatic thermodynamic process can be characterised by introducing a metric on the space of Gibbs states, in such a way that minimally-dissipating  protocols  correspond to geodesic trajectories.
  Here, we show how to  generalize this approach to open quantum systems by finding the thermodynamic metric associated to a given  Lindblad master equation. 
  The obtained metric can be understood as a perturbation over the background geometry of equilibrium Gibbs states, which is induced by the Kubo-Mori-Bogoliubov (KMB) inner product. 
  We illustrate this construction on  two paradigmatic examples:  an Ising chain and  a two-level system interacting with a bosonic bath with different spectral densities.
 
\end{abstract}

\section{Introduction}
A central task in finite-time thermodynamics is to design protocols that maximise the extracted work while minimising the dissipation during the process. In the slow driving regime, a powerful approach consists in equipping the space of thermodynamic states with a metric whose geodesics correspond to minimally dissipative processes. This geometrical construction was first developed in the 80s for macroscopic endoreversible thermodynamics in a series of seminal papers~\cite{Weinhold, Weinhold2,Schlogl1985, Salamon1, Salamon3, Nulton1, Andresen,diosi1996thermodynamic}, and more recently it was extended to the microscopic regime \cite{Crooks,Zulkowski,Feng2009,Sivak2012}, leading to several  applications in, e.g.,  molecular motors \cite{Sivak2016} and small-scale information processing~\cite{zulkowski2013optimal,Zulkowski2015}. 

While this  approach is well established for classical systems, the quantum regime has remained less explored. The geometry of quantum equilibrium Gibbs states has been characterised in~\cite{PetzB, Petz2000, Petz, PetzG, Balian}{;}  however the resulting metric does not take into account dynamical features of the dissipation, which are of crucial importance in finite-time protocols. For arbitrary out-of equilibrium evolutions, a notion of thermodynamic length has been put forward in~\cite{Deffner2010,Deffner2013}; yet, this approach requires full knowledge of the global unitary evolution, which makes it   difficult to apply in common situations where the size of the bath allows only for an effective description of the dynamics.   Finally, Kubo linear-response theory also allows for describing dissipation near equilibrium \cite{Campisi2012geometric,acconcia2015shortcuts,Ludovico2016adiabatic}, which in turn allows for defining a notion of thermodynamic metric \cite{acconcia2015shortcuts,Bonanca2018optimal}.

 The goal of this article is to 
 provide a general framework to construct a thermodynamic metric whenever the evolution of the system can be described by a Lindblad master equation {(see also~\cite{Zulkowski2015aa})}. 
%In particular, it appears that the metric obtained when one neglects the dynamics of the system has the role of a background geometry, which is then modified by the Drazin inverse of the Lindblad operator, encoding the equilibration timescales for different observables. 
The obtained metric can {be} expressed as the background equilibrium geometry of \cite{PetzB, Petz2000, Petz, PetzG}  acted upon by the Drazin inverse of the Lindblad operator, which encodes the different equilibration timescales of the dissipative dynamics. This geometrical approach is used to find minimally dissipating protocols of a slowly driven Ising chain in a transverse field, and of a qubit in contact with a bosonic bath with different spectral densities.

\section{{Dissipation in quantum systems}}
Before presenting the construction of the metric, we  here define the quantum correspondent of some important quantities of classical thermodynamics. A quantum system is described by its density matrix $\rho$ and its Hamiltonian $H$; if the state is given by the Gibbs ensemble we will use the notation $\rho \equiv \omega_\beta(H)$, where $\omega_{\beta}(H)=e^{-\beta H}/\Tr(e^{-\beta H})$ and $\beta$ is the inverse temperature of the surrounding  bath. {We will sometimes also use the shorthand notation $\omega:=\omega_{\beta}(H)$.} A functional of key importance is the non equilibrium free energy, which is defined by the formula $F(\rho, H) = \average{H}_\rho -\beta^{-1} S(\rho)$, where {we define the average as} $\average{A}_\rho =\Tr\sqrbra{A\rho}$, and we denote by $S(\rho)$ the von Neumann entropy of the state.

Direct calculations show that the non equilibrium free energy is connected to the equilibrium one by the equality
\begin{align}\label{eq:noneqFreeEnergy}
F(\rho, H) =F(\omega_\beta(H), H) +\beta^{-1} S(\rho||\omega_\beta(H))
\end{align}
where $ S(\rho||\omega_\beta(H))$ is the relative entropy. This quantity is positive definite and it  can be understood as a measure of how statistically different the current state is from the thermal one. Moreover, it corresponds to the extractable work from $\rho$  and, for this reason, it is sometimes referred to as availability.

We consider thermodynamic processes in which the Hamiltonian of a system in contact with a thermal bath is experimentally varied between two fixed endpoints, say $H_A$ and $H_B$. Each protocol is then defined by a curve $\gamma$ in the space of controllable parameters, and by its duration $T$. The work extracted is defined as 
\begin{align}\label{eq:work}
W =  -\int_\gamma \text{d}t\, \Tr\sqrbra{\rho_t \dot H_t}.
\end{align}
It is useful for what follows to rewrite the work in terms of the non equilibrium free energy.  Using the insight coming from~\eqref{eq:noneqFreeEnergy}, we integrate equation~\eqref{eq:work} by parts and obtain (see Appendix~\ref{app:Derivation1}):
\begin{align}\label{eq:workSplit}
W = -\Delta F_{\rm n.eq.} - \beta^{-1}\int_\gamma \text{d}t\, [-\partial_{\rho_t} S(\rho_{t} || \omega_\beta(H_t))],
\end{align}
where $\Delta F_{\rm n.eq.} $ is the difference in the non equilibrium free energy at the endpoints, and $ \partial_{\rho_t} S(\rho_t||\omega(H_{t}))$ is given by the limit
\begin{align}\label{eq:derivativeRelativeEntropy}
\lim_{\varepsilon\rightarrow 0} \frac{S(\rho_{t+\varepsilon} || \omega_\beta(H_t))-S(\rho_{t} || \omega_\beta(H_t))}{\varepsilon} .
\end{align}
The importance of equation~\eqref{eq:workSplit} is that it makes manifest that we can isolate a term which only depends on the endpoints of the protocol, and a contribution, $W_{\rm diss}(\gamma) = -(W+\Delta F_{\rm n.eq.})$, which accounts for the dissipation during the process and that explicitly depends on the path $\gamma$. The quantity inside the integral can be identified with the entropy production rate, which is guaranteed to be positive definite whenever one has Markovian dynamics (see Appendix~\ref{app:EntropyProd}). {This identification is also justified by noting that the heat supplied to the system can be rewritten as:
\begin{align}\label{eq:heatSplit}
	\beta \Delta Q = \Delta S - \int_\gamma \text{d}t\, [-\partial_{\rho_t} S(\rho_{t} || \omega_\beta(H_t))],
\end{align}
so that the integral accounts for the correction to the second law in the presence of dissipation (Appendix~\ref{app:Derivation1}).
}

%It should be noticed that 
%{We now focus on the slow driving limit, where the characteristic time scale of the dr}
{When the driving of $H_t$ is slow  (i.e. the characteristic timescale of the driving is longer than the relaxation timescales), }
%If the duration of the protocol is much bigger than the equilibration timescales of the system {and the protocol is sufficiently smooth}, 
it is sensible to assume that the entropy production rate will only depend on the base point $H_t$ and on the velocity $\dot H_t$. The key idea of the classical papers~\cite{Salamon1, Salamon3, Nulton1,Andresen} is then to reduce the task of finding the minimally dissipating path $\gamma^*$ to the problem of finding the geodesic of the metric induced by the entropy production rate on the thermodynamic space. We will now show how to apply this idea to open quantum systems.
% quantum regime.

\section{Metric structure in open quantum systems}
We assume that the dissipative dynamics is described by a time-dependent Lindblad equation,
\begin{align}
\dot{\rho}_t=\lind_t[\rho_t]
\label{eq:masterequation}
\end{align}
Since a Lindbladian master equation always leads to Markovian dynamics~\cite{Gorini}, this assumption guarantees the positivity of the entropy production rate. 
Furthermore, we assume that the Lindbladian operator has a unique zero eigenstate, given by the instantaneous Gibbs state ($\mathcal{L}_t[\omega_\beta(H_t)]=0$), and that all the other eigenvalues have strictly negative real part. %This type of  dynamics are known as relaxing or mixing~\cite{MixingLindbladian}, since 
These two assumptions are sufficient to ensure that given any initial conditions the system will thermalise to the instantenous Gibbs state of $H_t$,
\begin{align}
\lim_{\tau\rightarrow\infty} e^{\tau\lind_t} \rho = \omega_\beta(H_t).
\end{align}
{As a final remark, note that by assuming   $\mathcal{L}_t(\omega_\beta(H_t))=0$ we are neglecting 
non-adiabatic contributions to $\mathcal{L}_t$,  which is justified whenever the bath dynamics are fast compared to the driving rate of the system Hamiltonian \cite{Albash2012,Yamaguchi2017,Dann2018}. }

{For slow driving of $H_t$, }
%If the duration of the protocol is much bigger than the relaxation timescales of the system, 
we can assume that the state is always close to the equilibrium one: $\rho_t = \omega_\beta(H_t) +\delta\omega_t$. Plugging this expansion in equation~\eqref{eq:masterequation}, we obtain an equation for $\delta\omega_t$ as~\cite{CavinaSlow,Mandal2016,crooks2018drazin}:
\begin{align}
\norbra{ \lind_t - \frac{\text{d}}{\text{d}t}}[\delta\omega_t] = \frac{\text{d}}{\text{d}t}\omega_{\beta}(H_t).\label{eq:expandendMasterequation}
\end{align}
It is useful to express the derivative on the right hand side { using the derivative for the exponential of operators~\cite{Denes}:
\begin{align}
\frac{{\rm d} e^{-\beta H_t}}{{\rm d} t}=-\beta \int_0^1 {\rm d}s \hspace{1mm} e^{-\beta s H} \dot{H}_te^{-\beta (1-s) H},
\label{derexp}
\end{align}}%{Dyson's expansion for the exponential~\cite{Denes}},
obtaining $\dot \omega_\beta(H_t) = -\beta \J_{\omega_{\beta}(H_t)}[\dot H_t]$, where we defined the operator:
\begin{align}
\J_{\rho}[A]:=\int_0^1 \hspace{-0.5mm}{\rm d}s \hspace{1mm}   \rho^{1-s}\left(A - \Tr[\rho A]\id\right)\rho^{s}.
\label{eq:operatorJ}
\end{align}

In order to solve equation~\eqref{eq:expandendMasterequation} we need to introduce the Drazin inverse of the Lindbladian operator $\lind^+_t$ 
\begin{align}\label{eq:Drazin}
\lind^+_t[A]:=\int^\infty_0 \text{d}\nu \ e^{\nu\mathcal{L}_t}\big(\omega_\beta(H_t)\tr\sqrbra{A}-A\big).
\end{align}
which is the unique operator satisfying the three conditions (see~\cite{Boullion1971} and Appendix~\ref{app:DrazinInverse} for details): (i) commutation with the Lindbladian ($\lind_t\lind^+_t[A]=\lind^+_t\lind_t[A]=A-\omega_\beta(H_t)\tr\sqrbra{A}$), (ii) invariance of the thermal state ($\lind^+_t[\omega_\beta(H_t)]=0$) and (iii) tracelessness ($\tr\sqrbra{\lind^+_t[A]}=0$). 

An alternative expression of $\lind_t^+$ can be constructed in finite dimensions as follows~\cite{CavinaSlow}: Consider the space of traceless states, i.e., for any state $\rho_t$ we consider the traceless component given by $\delta\omega_t  := \rho_t- \omega_\beta(H_t)$. Thanks to the stationarity of $\omega_\beta(H_t)$, equation~\eqref{eq:masterequation} can be rewritten {in the instantaneous basis} as: {$ \delta\dot\omega_t = \Lambda_t[\delta\omega_t]$}, where $\Lambda_t$ is the projection of $\lind_t$ on the space of traceless states. Since the only zero eigenstate of the Lindbladian is traceful, $\Lambda_t$ is invertible on this subspace and we can identify $\lind_t^+ \equiv\Lambda_t^{-1}$, thanks to the uniqueness of the Drazin inverse. From this expression it is clear that $\lind_t^+$ encodes the thermalisation timescales of the system.

We can proceed to solve equation~\eqref{eq:expandendMasterequation}: multiplying both sides by $\lind^+_t$ and inverting $\norbra{ \id - \lind_t^+\frac{\text{d}}{\text{d}t}}$ we obtain
\begin{align}\label{eq:iterativeExpansion}
\rho_t &= \sum_{n=0}^\infty \norbra{\lind^+_t\frac{\text{d}}{\text{d}t}}^n \omega_\beta (H_t) =\nonumber\\
&=\omega_\beta (H_t) -\beta \lind^+_t  \hspace{-1mm}\left[ \mathbb{J}_{\omega_{\beta}(H_t)} [\dot H_{t}] \right] +\dots
\end{align}
In the slow driving limit each $\dot H_t$ is of order $\bigo{1/T}$ {(recall that $T$ is the total time of the process)}, so that we can truncate the perturbative expansion above at the second term, with corrections of order $\bigo{1/T^2}$. {More precisely, this approximation implicitly assumes that the trajectory is sufficiently smooth so that  derivatives of $H_t$ are always bounded, and that the duration of the protocol is much larger than the equilibration timescales encoded in $\lind_t^+$.}
%operator $\norbra{\lind^+_t\frac{\text{d}}{\text{d}t}}$ to be small during the whole protocol. This implies that in the presence of rapid variation of the Hamiltonian one has to consider a duration of the protocol big enough so to smear out this effects.}

\begin{figure}
	\centering
	\includegraphics[width=0.9\linewidth]{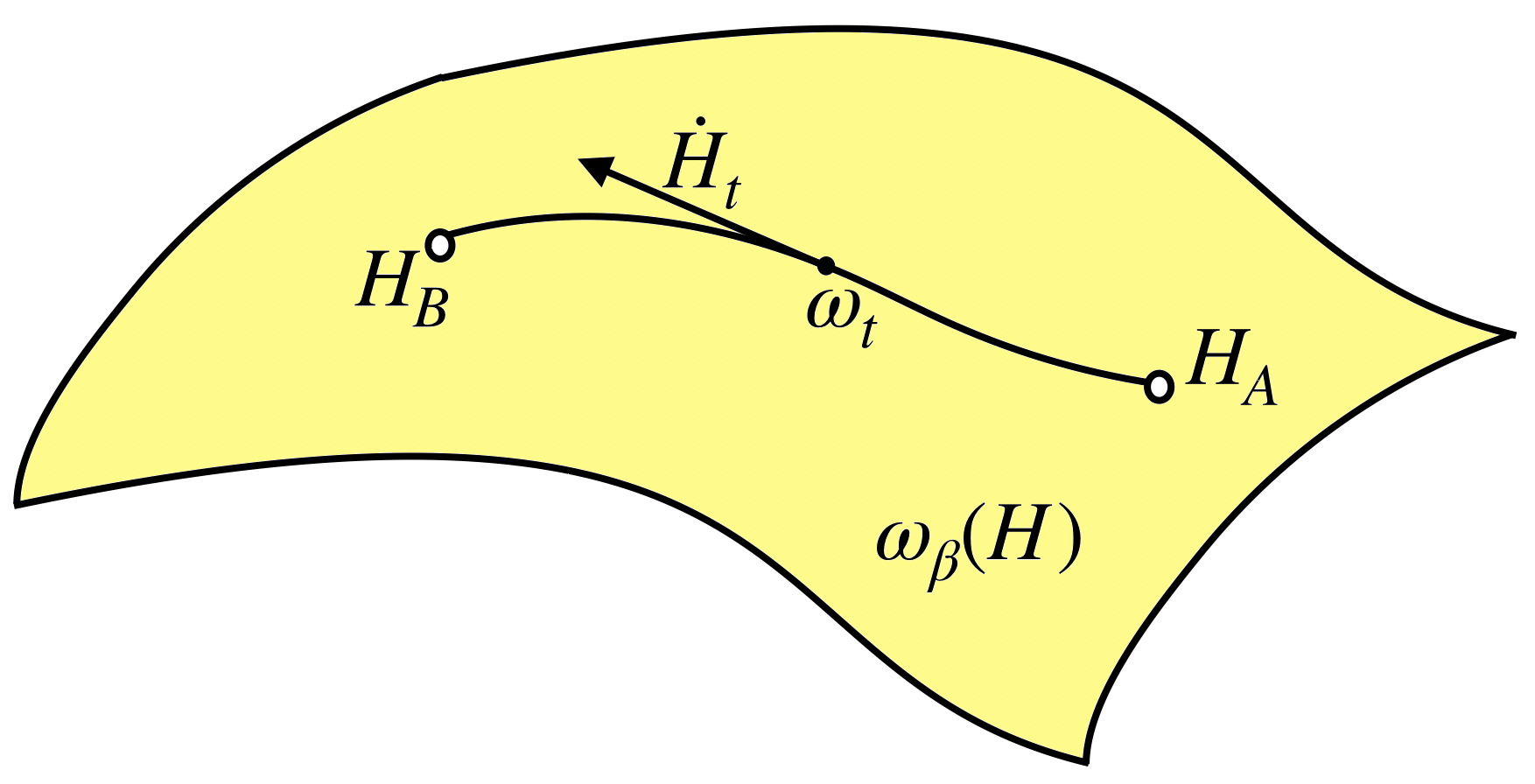}
	\caption{{A thermodynamic protocol in the slow driving regime naturally corresponds to a trajectory in the space of Gibbs states $\omega_{\beta}(H)$. The metric introduced lives in the tangent space of this manifold parametrized by $\omega_t := \omega_{\beta}(H_t)$}.}
	\label{fig:basepoint}
\end{figure}

Plugging expression~\eqref{eq:iterativeExpansion} into $ \partial_{\rho_t} S(\rho_t||\omega(H_{t}))$ gives the following formula for the dissipation (see Appendix~\ref{app:metricDerivation}):
\begin{align}\label{eq:workDiss}
W_{\rm diss} = -\beta\int_\gamma \text{d}t\, \Tr\sqrbra{\dot H_{t} \lind^+_t  \hspace{-1mm}\left[ \mathbb{J}_{\omega_{\beta}(H_t)} [\dot H_{t}] \right]}.
\end{align}
It should be noticed that this expression could be obtained directly by plugging the expansion of the state in the definition of the work~\eqref{eq:work}, but we preferred to keep the discussion independent of the particular expansion~\eqref{eq:iterativeExpansion}, so to emphasise the connection between $W_{\rm diss}$ and the entropy production rate.

At this point the introduction of a metric structure on the thermodynamic space is quite straightforward. Without loss of generality, we decompose the system Hamiltonian as ${H = \sum \lambda^i_t\, X_i}$, where $\curbra{\lambda^i_t} $ are the  {time-dependent externally controllable parameters}, and $\curbra{X_i}$ are the corresponding  observables. Then, equation~\eqref{eq:workDiss} can be rewritten as:
\begin{align}\label{eq:workMetric}
W_{\rm diss} = \beta\int_\gamma \text{d}t\, \dot\lambda^i_t \, m^\lind_{\omega_\beta(H_t)}(X_i, X_j ) \,\dot\lambda^j_t 
\end{align}
where we introduced the bilinear form:
\begin{align}\label{eq:metric}
m^\lind_\omega(A, B) = -\frac{1}{2}\Tr\sqrbra{A\, \lind^+_\omega  \hspace{-1mm}\left[ \mathbb{J}_{\omega} [B] \right] + B\, \lind^+_\omega  \hspace{-1mm}\left[ \mathbb{J}_{\omega} [A] \right]}
\end{align}
which is symmetric by construction, positive definite, thanks to the positivity of the entropy production rate, and it depends smoothly on the base point $\omega$. These are the defining properties of a metric. Note that, while we assumed  Lindblad master equation to ensure the positivity of \eqref{eq:metric}, the positivity also holds for more general maps~\cite{MllerHermes2017}, for which this approach can be naturally generalised.

Equation~\eqref{eq:workMetric} can be interpreted as the energy functional, or the action, of the curve $\gamma(\{\lambda_t\})$. This denomination can be understood thinking of {$W_{\rm diss}$} as the action of a system of point particles moving freely along $\gamma(\{\lambda_t\})$, with coordinate dependent mass encoded in $m^\lind_\omega$. Then, one can use Euler-Lagrange equations of motion~\cite{ONeil, Arnold1989} to construct curves in the space of parameters which minimize the dissipation at first order in $\bigo{1/T}$. This means that the geodesics induced by the metric~\eqref{eq:metric} correspond to minimally dissipative protocols {in the regime of slow processes}. %This will be discussed in Section \ref{Sec:MinDis}.

{
\subsection{Connection with classical results}
}

{The connection between the entropy production rate in the slowly driven regime and the geometry of Gibbs states has been studied for classical systems since the 70s~\cite{Weinhold, Weinhold2,Schlogl1985, Salamon1, Salamon3, Nulton1, Andresen}.} Historically, the formalism of thermodynamic length has been initially developed in the context of discrete perfectly thermalising  protocols, in which case the thermodynamic metric can be obtained as the second derivative of the partition function with respect to the control parameters~\cite{Crooks, Nulton1}. Let us now show how to recover this metric within our framework.

 %In order to better understand the connection between thermodynamic length for open quantum systems and the geometry of Gibbs states, it is useful to see how the construction presented here naturally generalises the classical formalism.

Let us assume that the parameters of the system are changed in discrete steps from $\curbra{\lambda^i_0} $ to $\curbra{\lambda^i_N}$. After each step we wait enough time for the system to perfectly equilibrate. In the quasistatic limit ($N\gg1$) we can use the approximation
\begin{align}
	H_{i+1 }= H_i + \tau \dot H_i + \bigo{\tau^2},
\end{align}
where $\tau$ is the time spent at each step, so that  the total time of the protocol is given by $T = N\tau$. Using the derivative  of the exponential \eqref{derexp} we also have up to order $\bigo{\tau^2}$ that:
\begin{align}\label{eq:discreteQuasistaticExpansion}
\omega_\beta (H_{i+1 })= \omega_\beta (H_{i}) - \tau\beta \mathbb{J}_{\omega_{\beta}(H_i)} [\dot H_{i}].
\end{align}
 
Comparing the expression just obtained with the expansion~\eqref{eq:iterativeExpansion}, it is straightforward to see that we could obtain the same result starting from the master equation,
\begin{align}
\dot{\rho}_{t}=\tau^{-1}(\omega_\beta (H_t)-\rho_t).
\label{eq:Gibbsmixing}
\end{align}
Moreover, the entropy production rate for discrete perfectly thermalising protocols coincides with the one coming from this fictitious evolution (see Appendix~\ref{app:discreteProcess}). This connection is a consequence of the fact that for discrete protocols details about the dynamics are disregarded, since at each step the state completely reaches equilibrium. This is translated into a master equation which has the least possible structure: in fact, the evolution~\eqref{eq:Gibbsmixing} for a fixed $t$ leads to an exponential relaxation for all observables with the same timescale~$\tau$.%We stress that the analogy betwe analogy is only valid in quasi-static processes, where $\rho_t$ is always close to $\omega_\beta (H_t)$.

In this context, we see that the metric~\eqref{eq:metric} reduces to the Kubo-Mori-Boguliobov (KMB) inner product
\begin{align}\label{eq:generalisedCov}
\hspace{-0.1cm}\frac{m^{\text{\tiny KMB}}_\omega(A, B)}{\tau} = {\rm{cov}}_\omega (A, B)= \Tr\sqrbra{A\,\mathbb{J}_{\omega} [B] },
\end{align} 
also known as generalised covariance in the context of linear response theory~\cite{Dyson1978, Roepstorff}. This quantity is directly related to the partition function by differentiation:
\begin{align}\label{eq:metricKMB}
	\beta^2 \,m^{\text{\tiny KMB}}_{i, j} = \frac{\partial^2}{\partial \lambda_i\partial \lambda_j}\log \mathcal{Z}_{H}	,
\end{align}
{where we indicate $m^{\text{\tiny KMB}}_{i, j}\equiv m^{\text{\tiny KMB}}_\omega(X_i, X_j)$, and $\mathcal{Z}_{H}$ is the partition function defined as: ${\mathcal{Z}_{H} = \Tr[{e^{-\beta H}}]}$.}

{In the classical case }the generalised covariance~\eqref{eq:generalisedCov} is substituted by the standard covariance {$\langle A B \rangle_{\omega}-\langle A \rangle_{\omega}\langle B \rangle_{\omega}$}~\cite{Crooks}.
%A similar expression was obtained in the classical paper~\cite{Crooks}, where the generalised covariance~\eqref{eq:generalisedCov} was substituted by the standard covariance $\langle A B \rangle_{\omega}-\langle A \rangle_{\omega} \langle B \rangle_{\omega}$. 
Indeed, when the protocol does not generate coherence between different energy levels, that is when $[H_t,\dot{H}_t]=0$, the former reduces to the latter. Hence, it is only in the case of non-commuting protocols, {meaning} $[H_t,\dot{H}_t]\neq 0$, that we expect qualitatively different results with respect to classical systems.

Similarly to the classical case, the metric~\eqref{eq:generalisedCov} can be interpreted as a quantum Fisher information matrix. The corresponding Riemannian structure has been studied both from the mathematical and the physical point of view~\cite{Balian, Petz2000, PetzB,guarnieri2019thermodynamics}. In particular,~\eqref{eq:metricKMB} {makes clear} the relation between phase transitions, statistical distinguishability and dissipation: in fact, close to a phase transition the eigenvalues of the Fisher information matrix corresponding to the order parameters diverge. Thanks to the extreme sensibility of the system, it is indeed easier to estimate the values of the external parameters {in this region}. In the same way, the abrupt changes in the system makes it more difficult to stay within the quasistatic limit, so that for any $N$ the dissipation will diverge close to the critical point.

In conclusion, the metric \eqref{eq:metric} extends the standard thermodynamic metric \eqref{eq:metricKMB} by including non-trivial relaxation dynamics through the Drazin inverse $\lind_t^+$. 
%At this point it  is clear that in the case of non-trivial relaxation dynamics  $\lind_t^+$ encodes the additional structure coming from the presence of different equilibration timescales.
 This modifies the underlying equilibrium geometry to account for the specific nature of the dissipative dynamics: since observables closer to their equilibrium value dissipate less when they are manipulated, the introduction of $\lind_t^+$ can be interpreted as a way to favour parameters which thermalise faster. A similar extension to non-trivial dynamics of the geometric framework for classical systems was obtained in~\cite{Sivak2012, Sivak2016} and the construction presented here constitutes the natural generalisation to the quantum regime. {In fact, if we consider the particular case in which the controllable observables satisfy the relaxation dynamics:
 	\begin{align}\label{eq:relDynamics}
\average{ \dot{X}_i}_{\rho_t}=  \tau_i^{-1}\norbra{\average{ X_i}_{\omega_t}-\average{ X_i}_{\rho_t}},
 	\end{align}
then the metric takes the simple form (see Appendix~\ref{app:metriccoordinates}):
\begin{align}\label{eq:metricSivak}
	m^\lind_\omega(X_i, X_j) = \frac{\tau_i+\tau_j}{2} \, m^{\text{\tiny KMB}}_{i, j},
\end{align}
in complete analogy with the classical result~\cite{Sivak2012}, which gives the thermodynamic length as the Hadamard product between a matrix encoding the equilibration timescales of the system and the Fisher information matrix. 
}
 %{As a final comment, we note that the metric~\eqref{eq:metricKMB}  can also be obtained in the context of collisional models \cite{Baumer2019imperfect}. } 

\section{Constructing minimally dissipating trajectories}\label{Sec:MinDis}

 We now discuss how to find minimially dissipative processes from the metric \eqref{eq:metric}, which is then applied to  an Ising chain in a transverse field, which equilibrates through \eqref{eq:Gibbsmixing}, and for a qubit in contact with a bosonic bath.

Before starting with the calculations some general remarks need to be made. First, it should be noticed that, in the case of partial control over the Hamiltonian  ${H = \sum \lambda^i_t\, X_i}$, the metric $m_{i,j}$ will be of the same dimension as the number of controllable parameters $\{ \lambda^i_t\}$ (see Appendix~\ref{app:metriccoordinates} for an explicit expression). The optimal trajectory can then be obtained by solving the geodesic equation:
\begin{align}\label{eq:geodesicEquation}
\ddot \lambda^i_t +	\Gamma^i_{j,k}\big|_{\lambda_t} \,\dot \lambda^j_t\, \dot \lambda^k_t = 0,
\end{align}
where $\Gamma$ denotes the Christoffel symbols, which can be computed from the formula:
\begin{align}\label{eq:chrisSymbols}
\hspace{-0.3cm}\Gamma^i_{j, k}|_{\lambda_t}  = \frac{1}{2} m^{i, l} \norbra{\partial_j m_{l, k} + \partial_k m_{j, l} - \partial_l m_{j, k}}|_{\lambda_t} ,
\end{align}
{where $m^{i, l}$ is the inverse of the metric, and we use the shorthand notation $\partial_i m_{j, k}|_{\lambda_t}\equiv (\partial m_{j, k}/\partial \lambda_i)|_{\lambda = \lambda_t}$.}

In this way, the machinery presented here provides an automatic method to obtain differential equations for the optimal trajectory directly in the space of parameters, which is usually of a dimension much smaller than the one of the full Hilbert space. Nonetheless, it is important to keep in mind that the computational cost still scales with the total dimension, where the most expensive operation is the computation of $m_{i,j}$ itself.

In the case of the  master equation~\eqref{eq:Gibbsmixing}, there are some important simplifications coming from the fact that the KMB metric is directly related to the partition function. In fact, one can obtain simply by differentiation:
\begin{align}
\beta^3 \,\Gamma^i_{j, k}|_{\lambda_t} =\frac{1}{2} m^{i, l}\norbra{ \frac{\partial^3}{\partial \lambda_j\partial \lambda_k\partial \lambda_l}\log \mathcal{Z}_{H}}.\label{eq:chrisKMB}
\end{align}
This makes  this approach  particularly simple when  an analytical expression for $\mathcal{Z}_{H}$ is known, as we will illustrate in the following example.

\subsection{Ising chain in a transverse field}
 
 We now consider the system to be given by an Ising chain in a transverse field, which equilibrates through \eqref{eq:Gibbsmixing}. The Hamiltonian of the system is given by:
\begin{align}\label{eq:hamiltonianIsing}
H_I  = -J \sum_{i=1}^n \norbra{\sigma_{i}^z \sigma_{i+1}^z + \,g\sigma^x_i }
\end{align}
where $J>0$ set a measure for the energy scale and $g$ is a dimensionless coupling parameter, which can be interpreted as an external magnetic field. For simplicity, in the following we will measure times in units of $\tau$ and energies in units of $J$. The spectrum of the system can be computed analytically for periodic boundary conditions through a Jordan-Wigner transformation, and in the thermodynamic limit the partition function is given by the integral expression (see e.g. ~\cite{sachdev2007quantum})
%, which maps the interacting Hamiltonian~\eqref{eq:hamiltonianIsing} to the free fermions model given by:
%\begin{align}
%H_I = \sum_k \varepsilon_k (\gamma_k^\dagger \gamma_k -\frac{1}{2})
%\end{align}
%where $\gamma_k^\dagger/\gamma_k$ are fermionic creation/annihilation operators and the corresponding energy is given by:
%\begin{align}
%\varepsilon_k = 2 \sqrt{1+g^2-2g \cos k}\,.
%\end{align}
%$k$ is evenly spaced in the interval $[0,2\pi]$. Due to the fermionic nature of the system, the partition function can be directly written as:
%\begin{align}
%\nonumber\mathcal{Z} &= \prod_k \sqrbra{\exp({ -\beta  \frac{\varepsilon_k }{2} })+\exp({ \beta  \frac{\varepsilon_k }{2} })} =\\
%&= \prod_k 2 \cosh\norbra{\beta\frac{\varepsilon_k }{2}}
%\end{align}
%In particular, in the thermodynamic limit we have the integral expression~\cite{sachdev2007quantum}:
\begin{align}\label{eq:logZ}
\lim_{n\rightarrow\infty} \frac{1}{N} \log \mathcal{Z} = \int_0^{2\pi} \text{d}k\, \log\sqrbra{2 \cosh\norbra{\beta\frac{\varepsilon_k }{2}}},
\end{align}
{where $\varepsilon_k $ is the eigenvalue corresponding to the momentum $k$, which reads:
\begin{align}
	\varepsilon_k = 2J \sqrt{1+g^2-2g \cos k}.
\end{align}
}

\begin{figure}
 	\centering
 	\includegraphics[width=1\linewidth]{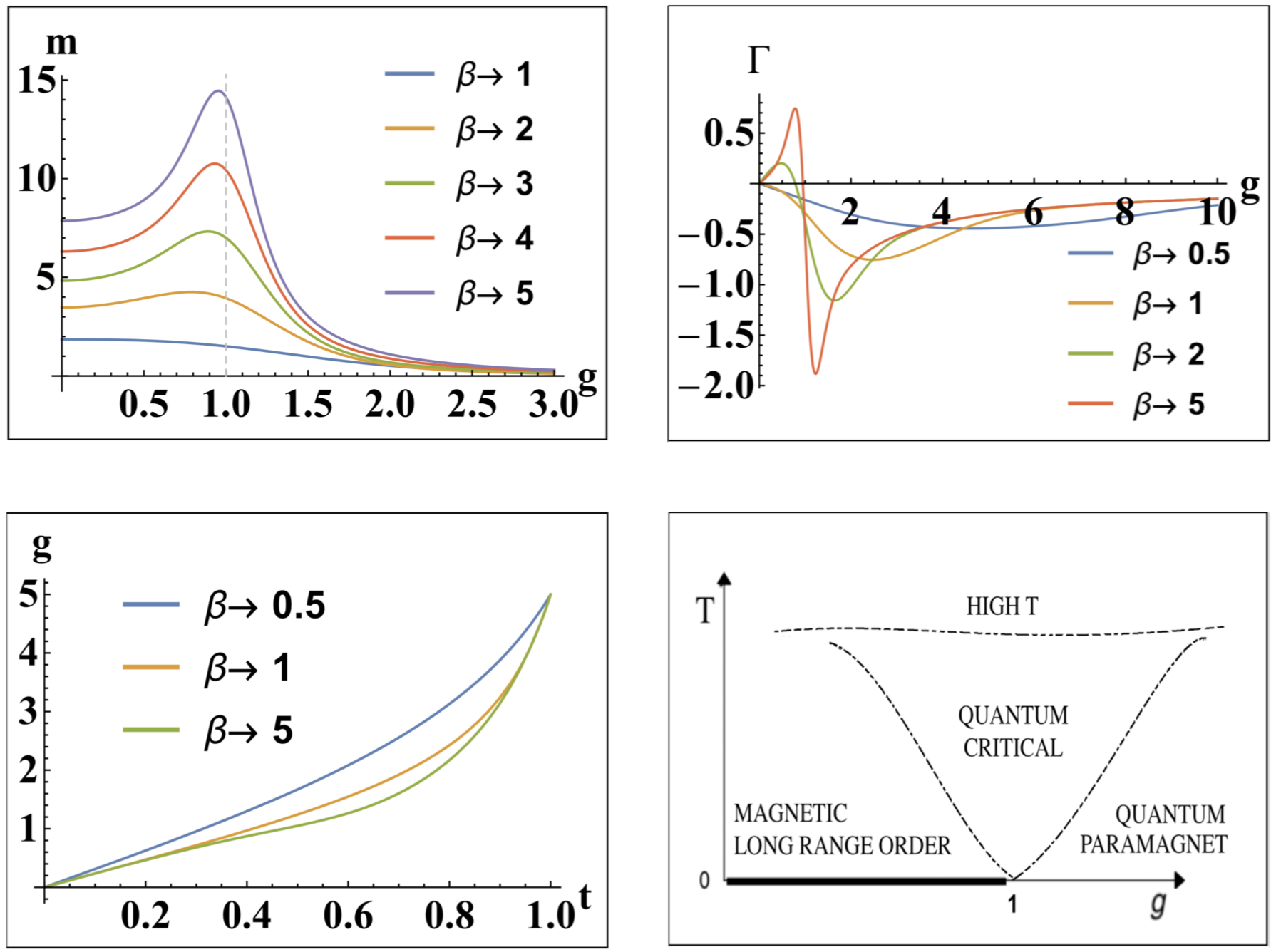}
 	\caption{The top two figures show the metric and the Christoffel symbol for the Ising chain at different temperatures as a function of the coupling $g$. On the bottom, on the left we present the behaviour of minimally dissipating trajectories defined by the boundary conditions $g(0)=0$ to $g(1)=5$; on the right, the phase diagram of the system.}
 	\label{fig:kmbgrid}
 \end{figure}

This system presents a phase transition at zero temperature and $g=1$ from an ordered ferromagnetic phase to a quantum paramagnetic phase. Moreover, at finite but low enough temperatures, there is a rich variety of different physical regimes (summarised in Fig.~\ref{fig:kmbgrid}), characterised by different scaling for the correlation length and the equilibration timescale of the system (see e.g. ~\cite{sachdev2007quantum} for details). %This behaviour is a signature of the zero temperature phase transition and it is thoroughly discussed in~\cite{sachdev2007quantum}. 
%, to which we refer for details about the derivation just presented.

We study here the case in which one has control only over $g$, so that the metric and the Christoffel symbols become a scalar, which we name $m$ and $\Gamma$, respectively. In particular, we see that it is sufficient to differentiate~\eqref{eq:logZ} with respect to $g$  twice, to obtain:
\begin{align}
&m(g) =\nonumber\\&\hspace{-1mm}\int_0^{2\pi}\hspace{-0.5mm} \text{d}k\, \norbra{\frac{\ddot\varepsilon_k }{2} \tanh\norbra{\beta\frac{\varepsilon_k }{2}} + \norbra{\frac{\dot\varepsilon_k }{2} {\rm sech}\norbra{\beta\frac{\varepsilon_k }{2}}}^2},
\end{align}
Similarly, we can compute $\Gamma$ using \eqref{eq:chrisKMB}. Both $m$ and $\Gamma$ are shown in Fig.~\ref{fig:kmbgrid} for different temperatures. Finally, we also compute numerically optimal thermodynamic processes through the geodesic equations \eqref{eq:geodesicEquation}. The results are also shown in Fig. \ref{fig:kmbgrid}. 

 Since the metric is connected with the free energy of the system, it is interesting to notice how the Riemannian structure is affected by the presence of a phase transition at zero temperature. Comparing the metric and, in particular, the Christoffel symbol with the phase diagram of the system, we can see that the change in behaviour of these geometric quantities retrace a change in the underlying physical properties. Additionally, note that $m$ increases close to $g=1$ as the temperature is decreased, illustrating how  the dissipation increases in the presence of a phase transition.  This behaviour reflects in the shape of the geodesics, as  $\dot{H}_t$ decreases close to the phase transition in order to compensate for the larger dissipation.

\begin{figure}
	\centering
	\includegraphics[width=1\linewidth]{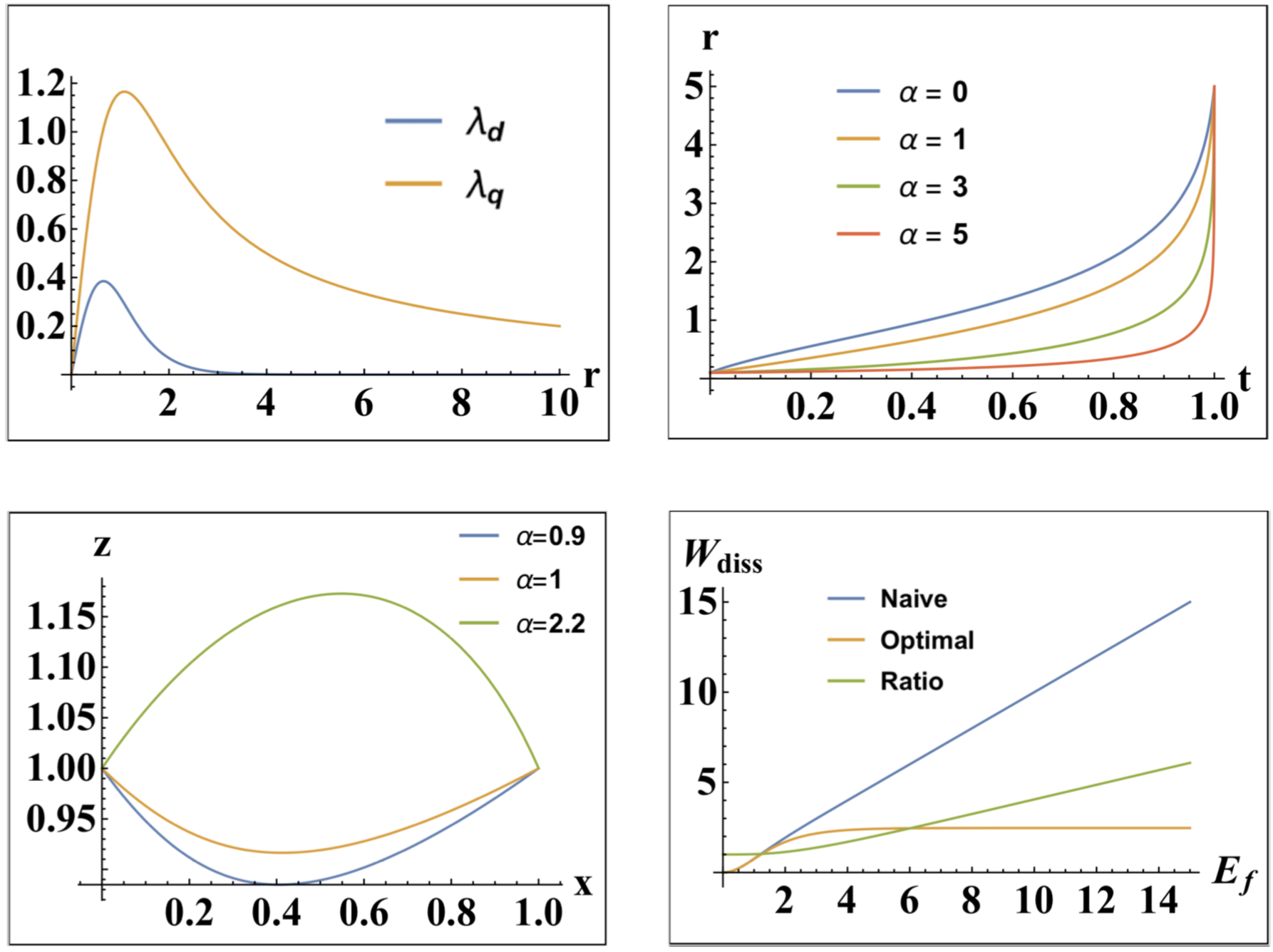}
	\caption{From left to right, from top to bottom: dependence of the classical and quantum component of the metric $\lambda_{d, q}$ on the energy spacing $r$; geodesics in the radial direction ($r(0)=0.1$, $r(1)=5$) {for constant $\theta$ and $\phi$}; geodesics {in Cartesian coordinates} for non commuting Hamiltonians ($\curbra{x(0),z(0)}=\curbra{0,1}$, $\curbra{x(1),z(1)}=\curbra{1,1}$) {and $y\equiv0$}; comparison of the {first order contribution to the} dissipation during a linear protocol with respect to the optimal one as a function of the energy spacing. The energies are measured in units of $\beta$.}
	\label{fig:qubitgrid}
\end{figure}
\subsection{Qubit in contact with a bosonic bath}
We  now move  to treat an example in which $\lind_t^+$ is non-trivial, a two level system in contact with a bosonic bath with spectral density $J(\omega)=\gamma_0 \omega^\alpha$. {The Lindbladian we use can be obtained through a microscopic derivation \cite{Petruccione}; more details and the particular form of the dynamics are given in appendix~\ref{app:LindbladianEvolution}.} The parameter $\alpha$ characterises the ohmicity of the environment, and we have that for $\alpha=1$, $\alpha>1$, $\alpha<1$, the bath is ohmic, superohmic and subohmic, respectively.  We assume full control on the Hamiltonian of the two-level system, which is  parametrised by spherical coordinates
\begin{align}
H = r\cos{\varphi}\sin{\theta} \,\hat \sigma_x+  r\sin{\varphi}\sin{\theta} \, \hat\sigma_y 
+  r\cos{ \theta} \, \hat\sigma_z,
\end{align}
where $(r, \theta, \varphi)$ are the control parameters. %Note that in this case the metric is 
%We assume full experimental control on the parameters $(r, \theta, \varphi)$. 
%We consider a bosonic bath with spectral density $J(\omega)=\gamma_0 \omega^\alpha$. 
%Again, the details about the derivation are provided in~\cite{SM}, and we will restrict here to the analysis of the results. 
Thanks to the convenient choice of coordinates, the metric takes the particularly simple form (see Appendix~\ref{app:LindbladianEvolution})
\begin{align}
m^{\lind} = \frac{1}{r^\alpha}{\rm{diag}}\curbra{\lambda_d,\, \lambda_q\, r^2,\, \lambda_q\, r^2\sin^2\theta},
\label{metrics}
\end{align}
where we can identify the expression of the Euclidean metric in spherical coordinates (${\rm diag}\{1,r^2,r^2 \sin^2(\theta)\}$), and the two eigenvalues are given by:
\begin{align}
\lambda_d = \frac{\tanh (r)} {\cosh^2(r)}, \hspace{1cm} \lambda_q =\frac{2 \tanh ^2(r)}{r}.
\end{align}
The existence of different eigenvalues for the radial direction and the solid angle reflects the physical fact that the system dissipates differently if only the energy spacing is moved, or if coherence between the two levels is created. Moreover, it is straightforward to verify that the ratio $\lambda_q/\lambda_d$ diverges for $r\rightarrow\infty$. This illustrates a general behaviour: the existence of exponentially more dissipative parameters in thermodynamic systems. In this case, there is a simple physical explanation: changing the energy spacing only affects an exponentially small fraction of the population, whereas the whole system has to be manipulated in order to create coherence. This example however illustrates the fact that inspecting the metric eigenvalues provides us with a powerful description of the physics of dissipation, which is  crucial  in more complex systems where a simple intuitive understanding is out of reach. % (e.g. in the Ising chain discussed above).

%Nonetheless, it is important to notice that the same conclusion could be reached simply by inspection of the eigenvalues of the metric, disregarding any physical intuition. In this way, it becomes possible to obtain a basic understanding of the physics of dissipation of more complex systems.

The ohmicity of the bath  contributes to the metric with a factor $r^{-\alpha}$, which is the overall dependence of the equilibration timescales on the energy spacing $r$. Owing to this dependence, optimal protocols will tend to spend more time in the region with low $r$. This effect is illustrated in the right top panel of Fig.~\ref{fig:qubitgrid}, where we construct optimal trajectories in the radial direction. As  expected, this effect is more pronounced for higher $\alpha$s.

Our results also allow for constructing optimal trajectories between non-commuting endpoints. Then we necessarily have that $[\dot{H}_t,H_t] \neq 0$, so that quantum coherence between energy levels is created along the protocol. Optimal non-commuting trajectories  are shown in the bottom left Figure of \ref{fig:qubitgrid}. Interestingly, we see that the ohmicity of the bath qualitatively changes the behaviour of the optimal trajectories.

In order to illustrate the improvement obtained by using geodesic trajectories, we compare them to a naive choice where the parameters are linearly modified in time. We consider  boundary conditions $H_A=0$ and $H_B=E_f \sigma_{z}$ (energy is measured in units of $\beta$). Restricting to trivial dynamics~\eqref{eq:Gibbsmixing},  which allows for a simple analytic solution, we find at  first order in $\bigo{1/T}$:
\begin{align}
&W_{\rm dis}^{\rm lin}=E_f \tanh{E_f},\\
&{W_{\rm dis}^{\rm KMB}}= \frac{1}{4}\left(\pi -2 \tan ^{-1}(\text{csch}(E_f ))\right)^2.
\end{align}
In the limit of  $E_f \gg 1$ the first term diverges linearly in $E_f$, while $W_{\rm dis}^{\text{\tiny KMB}} \approx \pi^2/4$. This difference makes clear how for big energy gaps or, equivalently, at low temperature, using a geodesic trajectory gives an increasingly bigger advantage, making the use of optimal trajectories particularly relevant in the quantum regime. These considerations are also relevant for experiments,  as current demonstrations of the Landauer principle rely on a linear increase of $E(t)$~\cite{berut2012experimental,Jun2014High,koski2014experimental,gaudenzi2018quantum}.

\section{Comparison to other approaches}

%In this section, we compare the approach put foward in this article with previous %It is important to point out how compares with work done in the past.
 Whereas the presented results are valid in the slow driving limit, exact solutions for minimising dissipation also exist in the literature~\cite{Seifert2007,esposito2010finite,Cavina2018a,Menczel2019}. In particular, optimal finite-time  protocols for two-level systems   have been treated in a variety of settings~\cite{esposito2010finite,deffner2014optimal,Cavina2018a,Cavina2018b,Menczel2019}. %single-level quantum dot~\cite{esposito2010finite}, for bosonic and fermionic baths in~\cite{Cavina,cavina2018variational}, and for a qubit in an optical cavity~\cite{deffner2014optimal}. 
Besides specific solvable systems, a general approach for minimising dissipation based on optimal control theory has been put forward in~\cite{Cavina2018a,Cavina2018b,Menczel2019}, which requires solving a system of $D$ non-linear differential equations, where $D$ is the Hilbert space dimension (see \cite{Cavina2018a} and Appendix~\ref{app:optControl}).

Conversely, the geometric approach   provides an approximation of the optimal solution (which becomes exact in the slow driving limit)  by a set of $d$ non-linear differential equations, where $d$ is the number of controllable parameters in the Hamiltonian (see Eq. \eqref{eq:geodesicEquation}). This makes this approach particularly useful in more complex systems, where $d\ll D$, as we have illustrated in the Ising chain example where $d=1$ and $D\rightarrow \infty$. This gives the geometric approach a wide range of applicability, which has also been illustrated in classical systems of different nature and complexity~\cite{Salamon1, Salamon3, Nulton1, Andresen,Sivak2016,zulkowski2013optimal,Zulkowski2015,bonancca2014optimal,
	Rotskoff2015,Gingrich2016,Rotskoff2017}.  It is however important to keep in mind that in general building the metric \eqref{eq:metric} requires diagonalising the Hamiltonian of the system of interest, %(or for the particular case of the master equation \eqref{eq:Gibbsmixing} finding the partition function), 
	as discussed in Appendix \ref{app:metriccoordinates}. {A class of dynamics where the computation of the metric notably simplifies is when the expected value of each externally controlled observable (i.e. $\langle X_i \rangle_{\rho_t})$) relaxes to its thermal expectation value with a well defined time scale $\tau_i$ as in \eqref{eq:relDynamics}; as in this case the corresponding metric \eqref{eq:metricSivak} can be directly computed  from the partition function.  }
	
Besides providing an efficient way of finding minimally dissipative paths, there are other advantages of working in the quasistatic regime:
\begin{enumerate}
\item By construction, the protocols developed are independent of the total time $T$. That is, one does not need to optimise the path for every duration of the protocol. 
\item When dealing with small systems, it is not only important to minimise the dissipated work, but also to minimise fluctuations. In the quasistatic limit, minimising dissipation guarantees the minimisation of {fluctuations} for classical systems \cite{Crooks}, and for commuting protocols in the quantum regime \cite{harrymatteo2019}.
\item The introduction of a Riemannian structure in the space of parameters provides an automatic way to infer the dissipative properties of the system, just by inspecting the eigenvalues of the metric.
\end{enumerate}
Finally,  in Appendix~\ref{app:optControl} we also qualitatively compare the geometric approach with  exact approaches for a two-level system finding good agreement between them even when the total time $T$ of the protocol is similar to  the time scale of relaxation, see also~\cite{Bonanca2018optimal}.

\section{Conclusions and outlook}
%We showed how the construction of a metric from the entropy production rate is a general and powerful method to obtain optimal paths in the quasistatic regime.
We developed a general framework to define a thermodynamic metric for systems that evolve under a Lindblad master equation, in such a way that geodesics  correspond to minimally-dissipative paths in the quasistatic regime.
The connection with the non-equilibrium free energy  makes it extremely versatile and an extension to more general settings seems to be foreseeable with only minor modifications. For example, it should be possible to extend our approach to generalised Gibbs ensembles~\cite{guryanova2016thermodynamics,lostaglio2017thermodynamic,halpern2016microcanonical,perarnau2016work} simply by redefining the non equilibrium free energy~\eqref{eq:noneqFreeEnergy} so to take into account the extra conserved quantities. We can even consider a dynamics which equilibrates to a non equilibrium steady states by simply substituting the thermal state in the definition~\eqref{eq:derivativeRelativeEntropy} of the entropy production rate with the appropriate steady state $\pi(H)$ ~\cite{Mandal2016}. {For non-equilibrium steady states, an interesting future direction is to consider relations between this geometric approach and quantum thermodynamic uncertainty relations \cite{guarnieri2019thermodynamics,Timpanaro2019thermodynamics}.}
Moreover, despite the fact that we assume a Lindblad master equation, and therefore weak coupling, thanks to the reaction-coordinate mapping, an extension to the strong coupling regime is possible~\cite{Gelin2009,Campisi2009Fluctuation,Hilt2011Hamiltonian,Gallego2014,
	gelbwaser2015strongly,strasberg2016nonequilibrium, Newman2017Performance,Marti}. In this context, since the spectral density of the bath is modified under the mapping, we expect that in the strong coupling regime there will be qualitatively changes in the behaviour of optimal trajectories, similarly to what we showed in Fig.~\ref{fig:qubitgrid}. Similarly, the same framework can also be applied to other types of thermalisation such as in collisional models \cite{Baumer2019imperfect}. %These are interesting extensions that are worth investigating in the future. 
{	Another interesting research direction is to investigate the connection between optimal thermodynamic processes and shortcuts to thermalisation \cite{Dann2019}}. 
Overall,  it seems that the introduction of a metric structure on the  space of thermodynamic states is not only natural, but it is worth of more investigation in the future.

~\\
\paragraph{Acknowledgements.} We thank G. Crooks, S. Deffner, M. Ueda, and  V. Cavina for insightful comments.  M.P.-L. acknowledges support from the Alexander von Humboldt Foundation. This project has received funding from the European Union’s Horizon 2020 research and innovation programme under the Marie Skłodowska-Curie grant agreement No 713729, and from Spanish MINECO (QIBEQI FIS2016-80773-P, Severo Ochoa SEV-2015-0522), Fundacio Cellex, Generalitat de Catalunya (SGR 1381 and CERCA Programme).
This research was also supported in part by the National Science Foundation under Grant No. NSF PHY-1748958. 

\bibliographystyle{unsrtnat}

\onecolumn\newpage
\appendix

\section{Non equilibrium free energy and work extraction}\label{app:Derivation1} 
In this section we give the details on how to derive equation~\eqref{eq:noneqFreeEnergy} and~\eqref{eq:workSplit} of the main text. First, it is straightforward to show, using the defining formula $F(\rho, H) = \average{H}_\rho -\beta^{-1} S(\rho)$, that the free energy of a thermal state $\omega_\beta(H)$ is given by
\begin{align}
F(\omega, H) = \average{H}_\rho -\beta^{-1} (\beta\average{H}_\rho + \log \mathcal{Z}_H) = -\beta^{-1} \log \mathcal{Z}_H
\end{align}
in complete analogy with classical thermodynamics. Then, equation~\eqref{eq:noneqFreeEnergy} is obtained by direct computation, rewriting the relative entropy as:
\begin{align}
\beta^{-1}S(\rho||\omega_\beta(H)) &= \beta^{-1}\,\Tr\sqrbra{\rho(\log\rho -\log\omega_\beta(H))}=\nonumber\\
&= \norbra{-\beta^{-1}\,S(\rho) +\average{H}_\rho} +\beta^{-1}\log\mathcal{Z}_H=\nonumber\\
&= F(\rho, H) - F(\omega_\beta(H), H),
\end{align}
which gives the result after rearranging the terms.
For what regards equation~\eqref{eq:workSplit}, it is useful to first give the expression of the total derivative of the non equilibrium free energy along a trajectory $(\rho_t, H_t)$. This can be decomposed in the two terms:
\begin{align}\label{eq:totalDerivative}
\frac{\text{d}}{\text{d}t} F(\rho_t, H_t) = (\Tr\sqrbra{\dot \rho_t H_t} + 	\beta^{-1}\,\frac{\text{d}}{\text{d}t} \Tr\sqrbra{\rho_t \log \rho_t}) + \Tr[\rho_t \dot H_t].
\end{align}
The last trace can be rewritten as:
\begin{align}
\Tr[\rho_t \dot H_t] = \lim_{\varepsilon\rightarrow 0} \frac{F(\rho_t, H_{t+\varepsilon})-F(\rho_t, H_{t})}{\varepsilon} =: \partial_{H_t} F(\rho_t, H_{t}).
\end{align}
For what regards the term inside the parenthesis, it should be first noticed that we can rewrite:
\begin{align}
\Tr\sqrbra{\dot \rho_t H_t} &= 	-\beta^{-1}\,(\Tr\sqrbra{\dot \rho_t \log e^{-\beta H_t}} -\log\mathcal{Z}_H\Tr\sqrbra{\dot \rho_t}) =\nonumber\\
&=-\beta^{-1}\,\Tr\sqrbra{\dot \rho_t \log \omega_\beta(H_t)},
\end{align}
where in the first line we use the fact that the derivative of a state is traceless. Then, we obtain for the whole expression:
\begin{align}
\frac{\text{d}}{\text{d}t} \Tr\sqrbra{\rho_t \log \rho_t} -\,\Tr\sqrbra{\dot \rho_t \log \omega_\beta(H_t)} = \lim_{\varepsilon\rightarrow 0} \frac{S(\rho_{t+\varepsilon} || \omega_\beta(H_t))-S(\rho_{t} || \omega_\beta(H_t))}{\varepsilon} =: \partial_{\rho_t} S(\rho_{t} || \omega_\beta(H_t)),
\end{align}
where we recognise the quantity defined in~\eqref{eq:derivativeRelativeEntropy}. In this way, equation~\eqref{eq:totalDerivative} can be expressed in the compact form:
\begin{align}
\frac{\text{d}}{\text{d}t} F(\rho_t, H_t) = \partial_{H_t} F(\rho_t, H_{t}) + \beta^{-1}\partial_{\rho_t} S(\rho_{t} || \omega_\beta(H_t))
\end{align}
This equality can be used to obtain equation~\eqref{eq:workSplit}. In fact, starting from the definition of the work in~\eqref{eq:work}, a simple integration by parts gives:
\begin{align}
W &=  -\int_\gamma \text{d}t\, \Tr\sqrbra{\rho_t \dot H_t} =-\int_\gamma \text{d}t\, \partial_{H_t} F(\rho_t, H_{t}) =\nonumber\\ 
&=  -\Delta F_{\rm n.eq.} -\beta^{-1} \int_\gamma \text{d}t\, [-\partial_{\rho_t} S(\rho_{t} || \omega_\beta(H_t))]\label{eq:work1}
\end{align}
This concludes the derivation of~\eqref{eq:workSplit}. 

{Finally, one can obtain equation~\eqref{eq:heatSplit} simply by using the first law of thermodynamics, plugging in the definition of $\Delta F_{\rm n.eq.}$:
\begin{align}
	\Delta Q &= \Delta U + W = \Delta U  -\Delta F_{\rm n.eq.} -\beta^{-1} \int_\gamma \text{d}t\, [-\partial_{\rho_t} S(\rho_{t} || \omega_\beta(H_t))]\nonumber\\
	&= \cancel{\Delta U} -\cancel{\Delta U} +\beta^{-1} \Delta S -\beta^{-1} \int_\gamma \text{d}t\, [-\partial_{\rho_t} S(\rho_{t} || \omega_\beta(H_t))].
	\label{eq:heat1}
\end{align}
Multiplying both sides by $\beta$, we obtain~\eqref{eq:heatSplit}.
}

\section{The entropy production rate}\label{app:EntropyProd}

We now study the main features of the entropy production rate. Equation~\eqref{eq:work1} suggests that the definition of the instantaneous entropy production rate is~\cite{Petruccione}:
\begin{align}
\dot\sigma(\rho_t):=& -\partial_{\rho_t} S(\rho_{t} || \omega_\beta(H_t)) = - \lim_{\varepsilon\rightarrow 0} \frac{S(\rho_{t+\varepsilon} || \omega_\beta(H_t))-S(\rho_{t} || \omega_\beta(H_t))}{\varepsilon}.
\label{eq:entropyProductionRate}
\end{align}
Before investigating the positivity of this quantity, it is useful to show a weaker equation for $\rho_t$ which are induced by CPTP maps {in the case in which there is no driving}. In particular, assuming the system {to be} initially uncorrelated with the environment ($\rho_{UNIV}(0) =\rho^S_0\otimes\rho^E_0$), and that the Gibbs state is stationary under the evolution, one has that:
\begin{align}
S(\rho_t || \omega_\beta(H)) &= S(\text{Tr}_E\sqrbra{U_t(\rho^S_0\otimes\rho^E_0) U_t^\dagger} || \text{Tr}_E\sqrbra{U_t(\omega_\beta(H)\otimes\rho^E_0) U_t^\dagger})\leq\nonumber\\
&\leq S(U_t(\rho^S_0\otimes\rho^E_0) U_t^\dagger || U_t(\omega_\beta(H)\otimes\rho^E_0) U_t^\dagger) =  S(\rho^S_0\otimes\rho^E_0|| \omega_\beta(H)\otimes\rho^E_0) =\nonumber\\
&= S(\rho^S_0 || \omega_\beta(H)) +\cancel{S(\rho^E_0 || \rho^E_0)},
\end{align}
where we used the fact that the relative entropy decreases when one traces out part of the system, together with its invariance under unitary evolution. 
Denoting with $V_t$ the dynamical map which bring the reduced density matrix of the system from the original state $\rho^S_0 $ to $\rho_t$, we can rewrite the equation as:
\begin{align}
S(\rho^S_0 || \omega_\beta(H)) - S(V_t\,\rho^S_0  || \omega_\beta(H)) \geq 0,
\end{align}
which ensures that on any finite time the integral of $-\partial_{\rho_t} S(\rho_{t} || \omega_\beta(H))$ is positive. Unfortunately, this result does not give any information about the monotonicity of the change of entropy: it is sufficient to consider an exactly recurrent system, with periodicity $T$, for which the the quantity $S(\rho^S_0 || \omega_\beta(H)) - S(\rho_{t} || \omega_\beta(H))$ will first increase, to only go back to zero at time $T$.

Nonetheless, when $V_t$ is a dynamical semigroup, meaning that for any $t,s \geq 0$ the identity $V_{t+s} = V_{t}V_{s}$ holds, one can further prove that:
\begin{align}
S(V_{t'}\, \rho || \omega_\beta(H)) = S(V_{t'-t}V_{t}\, \rho || \omega_\beta(H))\leq S(V_{t}\, \rho || \omega_\beta(H))\qquad \text{for } t'\geq t.\label{eq:positivityOfTheEntropy}
\end{align}
This result implies that the entropy production rate in \eqref{eq:entropyProductionRate} is positive. As it was stated above, this result is true in general only if the dynamics can be described by a dynamical semigroup, that is, if the evolution is Markovian. 

{The generalization to the case in which the system is slowly driven as considered here follows directly from the fact that } {we are dealing with adiabatic master equations for which $\mathcal{L}(\omega_\beta(H_t))=0$ $\forall t$}, {and hence we can restrict ourselves for each time $t$ to an infinitesimal region around $H_t$ and apply the same reasoning.}% {In fact, this extension is natural given that we are dealing with adiabatic master equations in which $\mathcal{L}(\omega_t)=0$ $\forall t$ \cite{Albash2012,Yamaguchi2017,Dann2018}.} 
%\textcolor{red}{In principle, adiabatic corrections to the master equation steady state might only complicate the exposition without qualitatively change the results. Add some references?}

\section{Drazin inverse of the Lindblad operator}\label{app:DrazinInverse}

Here we derive the integral expression for the Drazin inverse $\lind^+_t$ in the space of trace-class operators. We first recall the three conditions needed to define the inverse:
\begin{align}
\label{eq:B1}&\lind_t\lind^+_t[A]=\lind^+_t\lind_t[A]=A-\omega_\beta(H_t)\tr\sqrbra{A}; \\
\label{eq:B2}& \lind^+_t[\omega_\beta(H_t)]=0; \\
\label{eq:B3}& \tr\sqrbra{\lind^+_t[A]}=0.
\end{align}
We introduce the following trial solution 
\begin{align}\label{eq:Drazintest}
\Lambda^+_t[A]:=\int^\infty_0 \text{d}\nu \ e^{\nu\lind_t}\big(\omega_\beta(H_t)\tr\sqrbra{A}-A\big).
\end{align}
We first check~\eqref{eq:B2}, which gives
\begin{align}
\Lambda^+_t[\omega_\beta(H_t)]=\int^\infty_0 \text{d}\nu \ e^{\nu\lind_t}\big(\omega_\beta(H_t)\tr\sqrbra{\omega_\beta(H_t)}-\omega_\beta(H_t)\big)=0,
\end{align}
which follows from the normalisation of $\omega_\beta(H_t)$. For what regards~\eqref{eq:B3} we find
\begin{align}
\nonumber \tr\sqrbra{\Lambda^+_t[A]}&=\int^\infty_0 \text{d}\nu \ \tr\sqrbra{e^{\nu\lind_t}\big(\omega_\beta(H_t)\tr\sqrbra{A}-A\big)}= \\
&=\int^\infty_0 \text{d}\nu \ \big(\tr\sqrbra{\omega_\beta(H_t)}\tr\sqrbra{A}-\tr\sqrbra{A}\big)=0,
\end{align}
where we used the fact that the propagator $e^{\nu\lind_t}$ is trace-preserving. Finally, in order to show that~\eqref{eq:B1} holds, we first find the following:
\begin{align}
\nonumber\lind_t \Lambda^+_t[A]&=\int^\infty_0 \text{d}\nu \ \frac{d}{\text{d}\nu}e^{\nu\lind_t}\big(\omega_\beta(H_t)\tr\sqrbra{A}-A\big) =\int^{\nu=\infty}_{\nu=0} \ d\big(e^{\nu\lind_t}\big)\big(\omega_\beta(H_t)\tr\sqrbra{A}-A\big){=} \\
\nonumber&=A-\omega_\beta(H_t)\tr\sqrbra{A}+\lim_{\nu\to\infty} e^{\nu\lind_t}\big(\omega_\beta(H_t)\tr\sqrbra{A}-A\big) =A-\lim_{\nu\to\infty} e^{\nu\lind_t} A= \\
&=A-\omega_\beta(H_t)\tr\sqrbra{A},
\end{align}
where we used the fact that $\forall t$ $\lim_{\nu\to\infty}e^{ \nu \lind_t}[B]=\omega_\beta(H_t)$,
for any normalised operator $B$. Moreover, it also follows from the definition that
\begin{align}
\nonumber \Lambda^+_t\lind_t[A]&=\int^\infty_0 \text{d}\nu \ e^{\nu\lind_t}\big(\omega_\beta(H_t)\tr\sqrbra{\lind_t[A]}-\lind_t[A]\big)=-\int^\infty_0 \text{d}\nu \ e^{\nu\lind_t}\lind_t[A]= \\
\nonumber&=-\int^{\nu=\infty}_{\nu=0} \ d\big(e^{\nu\lind_t}\big)A =A-\lim_{\nu\to\infty} e^{\nu\lind_t} A= \\
&=A-\omega_\beta(H_t)\tr\sqrbra{A}.
\end{align}
We thus conclude that $\Lambda^+_t=\lind^+_t$. In fact, assume that there exists  another $\tilde\Lambda^+_t\neq\Lambda^+_t$ satisfying conditions~(37-39). Then, $\forall A$ we would have:
\begin{align}
\tilde\Lambda^+_t\sqrbra{A} \deq{\eqref{eq:B2}} \tilde\Lambda^+_t\sqrbra{A-\omega_\beta(H_t)\tr\sqrbra{A}} &\deq{\eqref{eq:B1}} \tilde\Lambda^+_t \sqrbra{\lind_{t}\Lambda^+_t\sqrbra{A}} =\nonumber\\&= \tilde\Lambda^+_t\lind_{t} \sqrbra{\Lambda^+_t\sqrbra{A}}\deq{\eqref{eq:B1}} \Lambda^+_t\sqrbra{A}-\omega_\beta(H_t)\tr\sqrbra{\Lambda^+_t\sqrbra{A}}\deq{\eqref{eq:B3}}  \Lambda^+_t\sqrbra{A},
\end{align}
which is in contradiction with the assumption that $\tilde\Lambda^+_t\neq\Lambda^+_t$. This completes the derivation of the integral expression~\eqref{eq:Drazin}.

\section{Metric in open quantum systems}\label{app:metricDerivation}
We show here how to obtain equation~\eqref{eq:workDiss} of the main text from the expression of the entropy production rate in~\eqref{eq:work1} and~\eqref{eq:entropyProductionRate}. Before starting, it is useful to introduce the derivative of the logarithm of a density matrix~\cite{Denes}:
\begin{align}\label{eq:D1}
\J_\rho^{-1}[\delta\rho] := \lim_{\varepsilon\rightarrow 0}\,\frac{\log (\rho+\varepsilon\delta\rho) - \log(\rho )}{\varepsilon} = \int_0^\infty \text{d}x\, (x+\rho)^{-1} \delta\rho\,(x+\rho)^{-1}.
\end{align}
It is worth pointing out that this operator is the inverse of $\J_\rho$ defined in the main text, when one restricts $\rho$ to density matrices.

We can now pass to first express the entropy production rate~\eqref{eq:entropyProductionRate} for a Lindbladian evolution, and then plug the quasistatic expansion of the state~\eqref{eq:iterativeExpansion}. First notice that we can split at first order $\dot \sigma$ as:
\begin{align}\label{eq:D2}
\dot\sigma(\rho_t) &=  - \lim_{\varepsilon\rightarrow 0}\, \norbra{\frac{\Tr\sqrbra{\rho_{t+\varepsilon} \log \rho_{t+\varepsilon} } - \Tr\sqrbra{\rho_{t} \log \rho_{t} }}{\varepsilon} \, + \frac{\Tr\sqrbra{\rho_{t+\varepsilon} \log \omega_\beta(H_t) } - \Tr\sqrbra{\rho_{t} \log \omega_\beta(H_t)}}{\varepsilon}}{=}\nonumber\\
&= -\Tr\sqrbra{\lind_t[\rho_{t}] (\log \rho_{t}  - \log \omega_\beta(H_t))} -\lim_{\varepsilon\rightarrow 0}\,\frac{\Tr\sqrbra{\rho_{t} \log \rho_{t+\varepsilon} } - \Tr\sqrbra{\rho_{t} \log \rho_{t} }}{\varepsilon}  .
\end{align}
where we used the Lindblad equation $\dot\rho_t = \lind_t[\rho_t]$. We will now show that the last term is actually zero. In fact, since for $\varepsilon\ll1$, $ \rho_{t+\varepsilon}\simeq \rho_t +\varepsilon \lind_t[\rho_t]+\dots$, we can expand the logarithm as:
\begin{align}
&\lim_{\varepsilon\rightarrow 0}\,\frac{\Tr\sqrbra{\rho_{t} \log \rho_{t+\varepsilon} } - \Tr\sqrbra{\rho_{t} \log \rho_{t} }}{\varepsilon} \simeq \lim_{\varepsilon\rightarrow 0}\,\frac{ \Tr\sqrbra{\rho_{t} \log \rho_{t} } +\varepsilon \Tr\sqrbra{\rho_{t}\, \J_{\rho_{t}}^{-1}[\lind_t[\rho_t]] }  - \Tr\sqrbra{\rho_{t} \log \rho_{t} }}{\varepsilon}  = \nonumber\\
&= \Tr\sqrbra{\rho_{t}\, \J_{\rho_{t}}^{-1}[\lind_t[\rho_t]] } = \Tr\sqrbra{\lind_t[\rho_t]} = 0,
\end{align}
where in the second line we used the fact that $\J^{-1}$ is self-adjoint with respect to the trace inner product, meaning that: $\Tr\sqrbra{A\J^{-1}[B]} = \Tr\sqrbra{\J^{-1}[A]B}$. This property can be inferred directly from the definition~\eqref{eq:D1} and the cyclicity of the trace. Moreover, it is just a matter of computation to see that $\J^{-1}_\rho[\rho] = \id$. Finally, in the last equality we used the fact that $\lind_t[\rho]$ is traceless.

We can now pass to give the quasi-static expansion of the entropy production rate. Recall that in this limit the state is given by:
\begin{align}
\rho_t & =\omega_\beta (H_t) -\beta \lind^+_t  \hspace{-1mm}\left[ \mathbb{J}_{\omega_{\beta}(H_t)} [\dot H_{t}] \right] +\dots
\end{align}
at first order in $\bigo{1/T}$. Plugging this expression in~\eqref{eq:D2}, we can obtain the result presented in the main text:
\begin{align}
\dot\sigma(\rho_t) &=- \Tr\sqrbra{\lind_t[\rho_{t}] (\log \rho_{t}  - \log \omega_\beta(H_t))}  = 	\nonumber\\
&= \beta\, \Tr\sqrbra{\lind_t[\rho_{t}] \, \J_{\omega_{\beta}(H_t)}^{-1}[\lind^+_t  \hspace{-1mm}\left[ \mathbb{J}_{\omega_{\beta}(H_t)} [\dot H_{t}] \right] ] }= 	\nonumber\\
&= -\beta^2\,\Tr\sqrbra{\mathbb{J}_{\omega_{\beta}(H_t)} [\dot H_{t}] \, \J_{\omega_{\beta}(H_t)}^{-1}[\lind^+_t  \hspace{-1mm}\left[ \mathbb{J}_{\omega_{\beta}(H_t)} [\dot H_{t}] \right] ] }=\nonumber\\
&= -\beta^2\,\Tr\sqrbra{\dot H_{t} \,\lind^+_t  \hspace{-1mm}\left[ \mathbb{J}_{\omega_{\beta}(H_t)} [\dot H_{t}] \right] }
\end{align}
where in the second line we used the expansion of the logarithm, in the third line we plugged in the expansion of the state and used the property of the Drazin inverse, and in the last line the fact that $\J^{-1}$ is self adjoint. This conclude the derivation of~(11).

\section{Work for discrete processes}
\label{app:discreteProcess}
The natural extension of the definition of extracted work~\eqref{eq:work} for processes in which the Hamiltonian is changed through a series of $N$ quenches is given by:
\begin{align}
	W = \sum_{i=0}^{N-1} \Tr\sqrbra{\rho_i (H_i - H_{i+1})},
\end{align}
where $H_i$ and $\rho_i$ are respectively the Hamiltonian of the system and its density matrix at the $i$-th step. Adding and subtracting at each step $S(\rho_i)$ and using the definition of nonequilibrium free energy, we can obtain a splitting analogous to~\eqref{eq:workSplit}:
\begin{align}
	W &= \sum_{i=0}^{N-1} (\Tr\sqrbra{\rho_i H_i} -\beta^{-1} S(\rho_i)) - (\Tr\sqrbra{\rho_i H_{i+1}} -  \beta^{-1} S(\rho_i)) = \nonumber\\
	&= \sum_{i=0}^{N-1} (F(\rho_i, H_i)-F(\rho_i, H_{i+1})) = \nonumber\\ 
	&= F(\rho_0, H_0)-F(\rho_N, H_{N})- \sum_{i=0}^{N-1} (F(\rho_i, H_{i+1})-F(\rho_{i+1}, H_{i+1})) =\nonumber\\
	&= -\Delta F_{\rm n.eq.} - \beta^{-1}\sum_{i=0}^{N-1} (S(\rho_{i}|| \omega_\beta(H_{i+1}))-S(\rho_{i+1}|| \omega_\beta(H_{i+1}))),
\end{align}
where in the last line we used the relation~\eqref{eq:noneqFreeEnergy}, which connects the nonequilibrium free energy with the equilibrium one. In this way, we can see that the entropy production rate for a discrete process is completely analogous to the one defined for a continuous evolution~\eqref{eq:derivativeRelativeEntropy}. 

If we now assume that at each step the system is allowed enough time to thermalise ($\rho_i \equiv \omega_\beta(H_{i})$) the dissipation simplifies to:
\begin{align}
W_{\rm diss} = \beta^{-1}\sum_{i=0}^{N-1} S(\omega_\beta(H_{i})|| \omega_\beta(H_{i+1})).
\end{align}

Before proceeding further, it is useful to point out that the expansion of the relative entropy is given by~\cite{Denes}:
\begin{align}
	S(\rho|| \rho +\varepsilon\sigma) = \varepsilon^2\ \Tr\sqrbra{\sigma \J^{-1}_\rho[\sigma]} +\bigo{\varepsilon^3}.
\end{align}
Then, if we assume to be in the quasistatic regime, so that we can use the expansion~\eqref{eq:discreteQuasistaticExpansion}, $W_{\rm diss}$ can be rewritten up to order $\bigo{\tau^2}$ as:
\begin{align}
W_{\rm diss} &= \beta^{-1}\sum_{i=0}^{N-1} S(\omega_\beta(H_{i})|| \omega_\beta(H_{i}) - \tau\beta \mathbb{J}_{\omega_{\beta}(H_{i})} [\dot H_{i+1}])  =\nonumber\\
&= \beta \tau^2\sum_{i=0}^{N-1}  \Tr\sqrbra{\mathbb{J}_{\omega_{\beta}(H_{i})} [\dot H_{i+1}] \J^{-1}_{\omega_{\beta}(H_{i})}[\mathbb{J}_{\omega_{\beta}(H_{i})} [\dot H_{i+1}]]}=\nonumber\\
&= \beta\tau \int_\gamma \text{d}t\, \Tr\sqrbra{\dot H_{t}\,  \mathbb{J}_{\omega_{\beta}(H_t)} [\dot H_{t}]} +\bigo{\tau^2},
\end{align}
where in the last line we used the definition of Riemann sum to pass from the discrete sum to the integral expression. As it can be seen, a formally equivalent result could be obtained from the fictitious master equation~\eqref{eq:Gibbsmixing}.

\section{Expression of the metric in coordinates}
\label{app:metriccoordinates}

In this section we explain how to get an explicit expression of the metric for general Lindbladian evolutions. As a preliminary remark, it should be notice that any operator $A$ can be treated as a vector in a linear space via the identification:
\begin{align}\label{eq:E1}
A =\sum_{l, m} A_{lm} |l\rangle \langle m| \longrightarrow {\ket{A}} =\sum_{lm} A_{lm} |lm\rangle,
\end{align}
where $|i\rangle$ is some orthogonal basis. In particular, since any Hermitian operator can be expressed as a linear combination of:
\begin{align}
\Delta_l = \ket{l}\bra{l},\hspace{1cm}\Sigma^x_{lm} =\frac{1}{\sqrt{2}} (\ket{l}\bra{m} + \ket{m}\bra{l}), \hspace{1cm} \Sigma^y_{lm} = \frac{i}{\sqrt{2}}(\ket{l}\bra{m} - \ket{m}\bra{l}) ,
\end{align}
we can interpret these operators as an orthonormal basis of the corresponding real vector space. Moreover, we can lift the Hilbert-Schmidt inner product to define a scalar product
\begin{align}
\braket{A}{B} = \Tr{[A^\dagger B]}.
\end{align}
This construction can be used to rewrite any linear superoperator as a $d^2\cdot d^2$, matrix, where $d$ is the dimension of the original Hilbert space.

Using this identification, we can give an explicit expression for the Drazin inverse of a mixing  Lindbladian. As it was said in the main text, Lindbladians of this type have a unique zero eigenvector, which we will denote by $\ket{\omega_{\beta}(H)}$, and all the other eigenvalues $\lambda_\alpha$ (corresponding to the eigenvectors $\ket{\alpha}$) have negative real part. Moreover, since $\lind_\omega$ is trace preserving, all the corresponding operators $\alpha$ must be traceless. We can give an explicit expression of $\lind^+_\omega$ as:
\begin{align}\label{eq:E4}
\lind^+_\omega&= \sum_{\alpha}(\lind^+_\omega)_\alpha \ket{\alpha} \bra{\alpha}= -\sum_{\alpha}\norbra{\int^\infty_0 \text{d}\nu \ \bra{\alpha} e^{\nu\lind}\ket{\alpha}} \ket{\alpha}\bra{\alpha}=\nonumber \\
&= -\sum_{\alpha}\int^\infty_0 \text{d}\nu \ e^{\nu\lambda_\alpha}\ket{\alpha}\bra{\alpha} = \sum_{\alpha}\frac{1}{\lambda_\alpha}\ket{\alpha} \bra{\alpha}
\end{align}
This explicit construction is a further proof of the fact that the Drazin inverse simply corresponds to the usual inverse restricted to the traceless subspace.

We can now pass to the explicit computation of the metric. For simplicity, we choose as a basis on the original Hilbert space the eigenbasis $\ket{i}$ of $H$. The only non-zero matrix element of $\J_{\omega_\beta(H)}$ in the basis  given by the identification~\eqref{eq:E1} can be explicitly evaluated as:
\begin{align}\label{eq:E5}
(\J_{\omega_\beta(H)})^{ij}_{lm}&= \frac{1}{\mathcal{Z}_H}
\int_0^1 \hspace{-0.5mm}{\rm d}s \hspace{1mm}  e^{-(1-s)\beta \epsilon_i}(\delta^i_l\delta^j_m-\frac{e^{-\beta \epsilon_l}}{\mathcal{Z}_H}\delta^i_j\delta^l_m)e^{-s\beta \epsilon_j}=\nonumber\\
&= \norbra{\frac{1}{ \mathcal{Z}_H}\frac{e^{-\beta \epsilon_i} -e^{-\beta \epsilon_j}}{ \beta\epsilon_j -\beta\epsilon_i}}\,\delta^i_l\delta^j_m -\norbra{\frac{e^{-\beta (\epsilon_i+\epsilon_l)}}{\mathcal{Z}_H^2}}\delta^i_j\delta^l_m.
\end{align}

Moreover, since we can rewrite the eigenbasis of $\lind_\omega$ as:
\begin{align}
\ket{\alpha} = u_{\alpha, ij} \ket{i j},
\end{align}
where $u$ is a unitary matrix, $\lind^+_\omega$ is expressed in this basis simply by:
\begin{align}\label{eq:E7}
\lind^+_\omega = \sum_{\alpha}\frac{1}{\lambda_\alpha}\ket{\alpha} \bra{\alpha} = \sum_{\alpha, i, j}  \frac{u_{ij, \alpha}^* u_{\alpha, lm} } {\lambda_\alpha}\ket{ij} \bra{lm}.
\end{align}
Now, assuming control only on a set of observables $\curbra{X^{(i)}}$, we can then finally compute the metric entries as:
\begin{align}\label{eq:E8}
(m^\lind_\omega)_{i,j} &= m^\lind_\omega(X^{(i)}, X^{(j)})=\nonumber\\
&= -\frac{1}{2}\sum_{lm,kn} X^{(i)}_{lm} X^{(j)}_{kn} ((\lind^+_\omega)^{ml}_{xy}(\J_{\omega_\beta(H)})^{xy}_{kn}+(\J_{\omega_\beta(H)})^{ml}_{xy}(\lind^+_\omega)^{xy}_{kn}).
\end{align}
However cumbersome at a first look, this expression corresponds to a simple matrix product of the operator specified in coordinates in~\eqref{eq:E5} and~\eqref{eq:E7}. In this way, one gets a matrix expression of $m^\lind_\omega$ of the same dimension as the one of controllable parameters. 

{We can now derive equation~\eqref{eq:metricSivak} in the main text.  Considering an evolution equation as in~\eqref{eq:relDynamics}, the Lindbladian can be expressed as:
\begin{align}
	\lind_\omega = -\sum_i \tau_i^{-1}\ket{X_i-\average{X_i }_\omega}\bra{X_i-\average{X_i }_\omega}.
\end{align}
Then, we can use equation~\eqref{eq:E4} to find the Drazin inverse of $\lind_\omega$, and plugging this expression in~\eqref{eq:E8} simply gives:
\begin{align}
		m^\lind_\omega(X_i, X_j) &=\frac{1}{2}\sum_{lm,kn} X^{(i)}_{lm} X^{(j)}_{kn} (\tau_i\, (\J_{\omega_\beta(H)})^{ml}_{kn} + \tau_j\, (\J_{\omega_\beta(H)})^{ml}_{kn}) = \nonumber\\
		&= \frac{\tau_i+\tau_j}{2} \, m^{\text{\tiny KMB}}_{\omega}(X_i, X_j),
\end{align}
which proves the claim.
}

\section{Two level system coupled to a bosonic bath}\label{app:LindbladianEvolution}

In this section the details about the Lindblad master equation of a qubit in contact with a bosonic bath. We {preliminary} choose the basis of the Hilbert space in such a way that the Hamiltonian can be rewritten as:
\begin{align}
H = \frac{1}{2}\,r \hat\sigma_z.
\end{align}
The dynamics induced {in the interaction picture} by a bosonic bath with spectral density $J(r)\propto r^{\alpha}$ is described by the master equation \cite{Petruccione}:
\begin{align}
\dot\rho_t &= \gamma_r(P_r+1)\left(\hat\sigma_-\rho_t\hat\sigma_+ - \frac{1}{2}\curbra{\hat\sigma_+\hat\sigma_-,\rho_t}\right) +\gamma_r P_r\left(\hat\sigma_+\rho_t\hat\sigma_- - \frac{1}{2}\curbra{\hat\sigma_-\hat\sigma_+,\rho_t}\right),
\label{eq:lindbladian}
\end{align}
where $\gamma_r$ and $P_r$ are given by:
\begin{align}
\gamma_r = \tilde\gamma_0 r^{\alpha}\hspace{2cm}
P_r = \frac{1}{e^{2 \beta r}-1}.
\end{align}
For simplicity, the proper equilibration timescale $\tilde\gamma_0^{-1}$ is assumed to be one. Additionally, for bookkeeping reasons it is useful to define the quantity $\Gamma_r = \gamma_r(2P_r+1)$. 
As it was argued in the main text, we can divide the density matrix of the system in a traceful and traceless component $\rho = \omega_\beta(H)+ \delta\omega$, so that the Lindblad equation can be rewritten as $\dot \rho_t  = \Lambda_t[\delta\omega_t]$. In this simple case, we can use the Pauli matrices to parametrize the traceless space:
\begin{align}
\begin{pmatrix}
z & x+iy\\ 
x-iy &-z 
\end{pmatrix} \xrightarrow{\qquad\qquad} (x, y, z).
\label{eq:stokesCoordinates}
\end{align}
and plugging this expansion into \eqref{eq:lindbladian} we obtain:
\begin{align}
\begin{pmatrix}
\dot x(t)\\
\dot y(t)\\
\dot z(t)
\end{pmatrix} = \Lambda_r\norbra{\rho(t) - {\omega_\beta(H)}} = \begin{pmatrix}
-\frac{\Gamma_r}{2} & 0 & 0\\
0 & -\frac{\Gamma_r}{2} & 0\\
0 & 0 & -\Gamma_r 
\end{pmatrix} 
\begin{pmatrix}
x(t)\\
y(t)\\
z(t) 
\end{pmatrix},
\end{align}
In this way we obtained a definition for $\Lambda_r$, and consequently for its Drazin inverse. {Assuming that the dissipative dynamics only depends on the energy spacing of the Hamiltonian, and not on the unit vector $\hat n = (\hat x, \hat y, \hat z)$, the Drazin inverse for a generic Hamiltonian is simply obtained by a change of basis. More explicitly we have:
\begin{align}
	H(x, y, z) = \frac{1}{2}r\, U \hat \sigma_z U^\dagger\hspace{0.5cm}\longrightarrow \hspace{0.5cm} \Lambda^{-1}_{(x,y,z)} = U  \Lambda^{-1}_r U^\dagger.
\end{align}
}

We now pass to quickly show how to obtain the metric in this case starting from the parametrisation in spherical coordinates:
\begin{align}
H = r\cos{\varphi}\sin{\theta} \,\hat \sigma_x+  r\sin{\varphi}\sin{\theta} \, \hat\sigma_y 
+  r\cos{ \theta} \, \hat\sigma_z.
\end{align}
Using the parametrisation~\eqref{eq:stokesCoordinates} the metric could be directly computed by plugging the Pauli matrices in equation~\eqref{eq:workMetric}. In spherical coordinates, though, one needs to first find the basis of the tangent space induced by the parametrisation, which can be computed by simple differentiation of the coordinate chart:
\begin{align}\label{eq:G7}
\left\{\begin{array}{l}
\partial_r = \cos{\varphi}\sin{\theta} \, \hat\sigma_x+  \sin{\varphi}\sin{\theta} \, \hat\sigma_y +\cos{\theta} \, \hat\sigma_z \\
\partial_\theta = r\cos{\varphi}\cos{\theta} \, \hat\sigma_x+  r\sin{\varphi}\cos{\theta} \, \hat\sigma_y - r\sin{ \theta} \, \hat\sigma_z \\
\partial_\varphi =  - r\sin{\varphi}\sin{\theta} \, \hat\sigma_x+  r\cos{\varphi}\sin{\theta} \, \hat\sigma_y.
\end{array}\right.
\end{align}

Then, the matrix form of the metric can be obtained directly from~\eqref{eq:E8}, using as {$\curbra{X^{(i)}}$} the observables~\eqref{eq:G7} {or, equivalently:
\begin{align}
	m_{i,\, j} = m^\lind_\omega(\partial_i, \partial_j) = -\frac{1}{2}\Tr\sqrbra{\lind^+_\omega  \hspace{-1mm}\left[ \mathbb{J}_{\omega} [\partial_j] \right]{\partial_i} + \lind^+_\omega  \hspace{-1mm}\left[ \mathbb{J}_{\omega} [\partial_i] \right]\partial_j}.
\end{align}
}
\section{Comparison with optimal control optimisation} \label{app:optControl}

The geometric approach provides an approximate solution which becomes exact in the limit of slow driving. In this appendix we compare it with the exact  solutions reported in \cite{esposito2010finite,cavina2018variational}.  First, we compare  in general terms the present  method to the  approach  presented in \cite{cavina2018variational} to minimise dissipation in open quantum systems, discussing how complex the two optimisation procedures are.  Then, we focus on a specific protocol involving a driven qubit and compare both approaches quantitatively. 
%We first compare in general terms the two methods,  to later do it in quantitative terms for a particular protocol involving a driven qubit. 

\begin{figure}
	\includegraphics[width= 0.4 \linewidth]{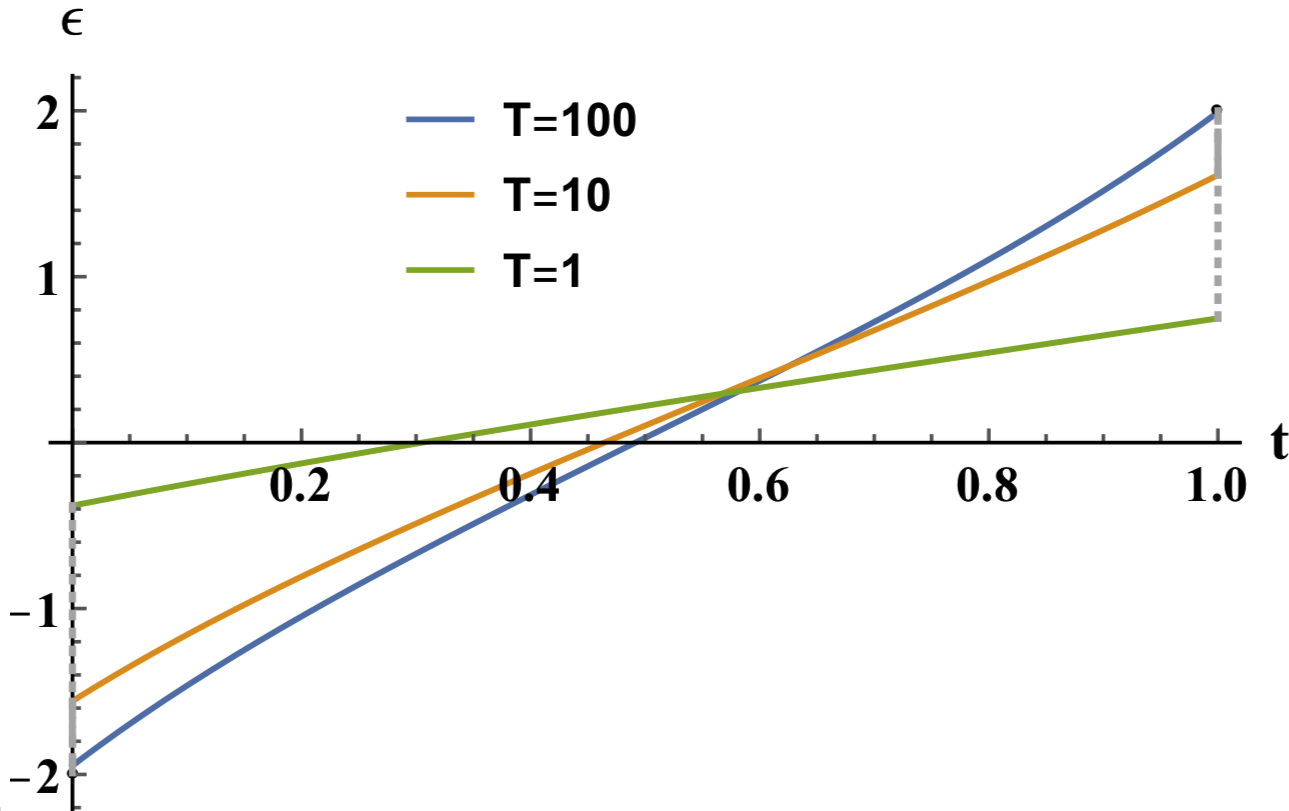}\hspace{1cm}
	\includegraphics[width= 0.4 \linewidth]{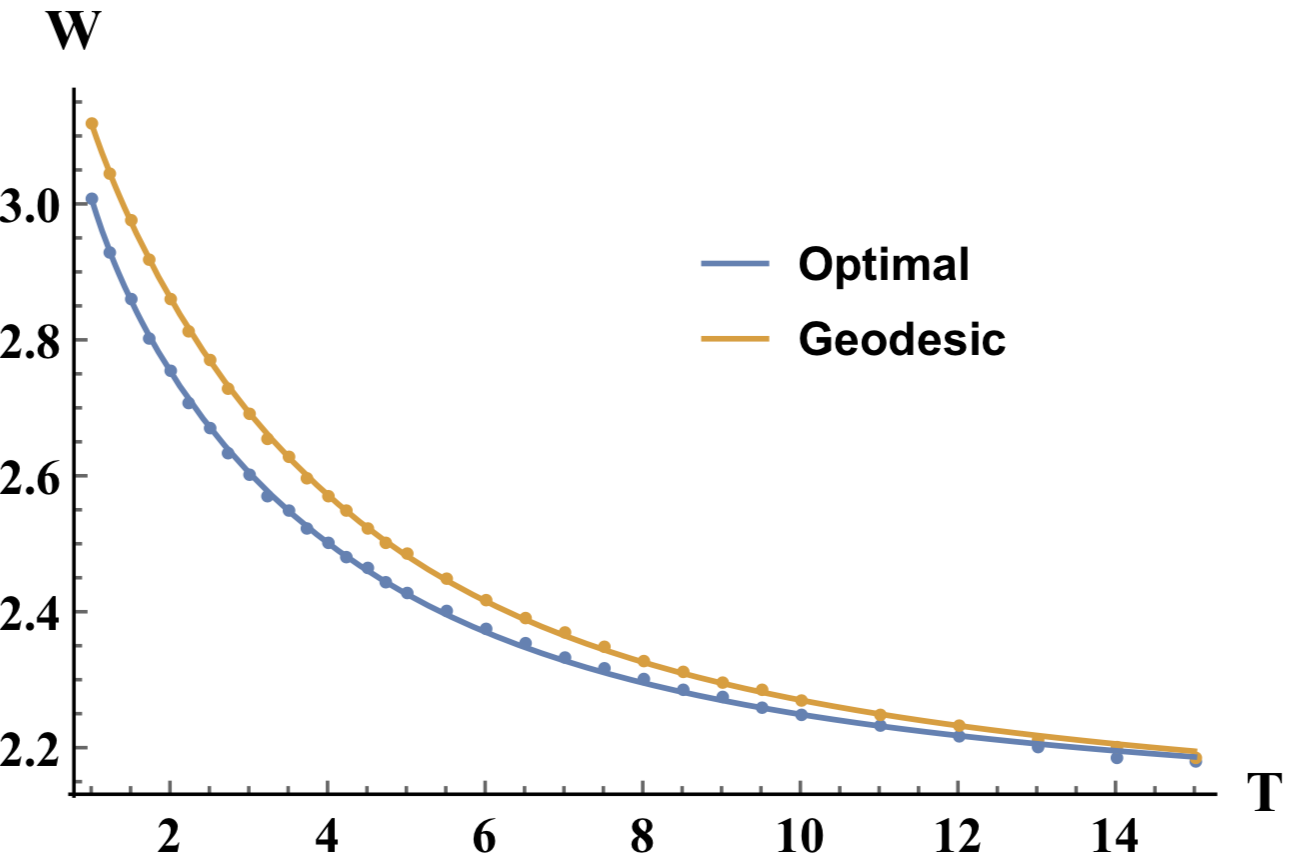}
	\caption{On the left hand side of the picture optimal protocols between $\epsilon(0) = -2 \beta$ and $\epsilon(1) = 2 \beta$ are presented for different times. These results are obtained with the technique presented in \cite{esposito2010finite}.  As it can be seen, for short times the trajectories present a quench at the beginning and at the end of the protocol. In the large time limit, instead, the optimal protocols become equivalent to the one obtained with the thermodynamic length. This behaviour is illustrated on the left hand side, where we plot the work required for trajectories obtained by exact minimisation or by solving the geodesic equation in function of time.}
	\label{fig:Comparison}
\end{figure}

The main idea of \cite{cavina2018variational} is  to use the Pontryagin’s minimum principle (PMP) to minimise the dissipation given a general Lindbladian master equation. In particular, using the decomposition of the system Hamiltonian ${H = \sum \lambda^i_t\, X_i}$, we denote by $d$ the number of externally controllable parameters and by $D$ the dimension of the Hilbert space. Then, following \cite{cavina2018variational},  one needs to solve a differential system of $2D$ equations  (see Eq. (16) in  \cite{cavina2018variational}), constraint by $d$ algebraic equations. This should be contrasted to the geometric approach reported here, which only requires the solution of $d$ differential equations (see Eqs.~\eqref{eq:metric},~\eqref{eq:geodesicEquation} and~\eqref{eq:chrisSymbols} in the main text). In general,  in the optimal control theory  there is no systematic way to reduce the number of differential equations from $2D$ to $d$. This is possible in particular systems, e.g., when the problem can be linearised; however, in thermodynamics the control parameters usually appear both in $H$ and $e^{-\beta H}$, making such a linearisation not possible. Hence,  while the complexity of the solution will of course depend on the specific system under considerations, it seems reasonable to say that the geometric approach will be  considerably simpler  in mesoscopic/many-body systems, in which usually $d\ll D$. An extreme case is shown for the Ising chain in the main text, where $d=1$ and $D \rightarrow \infty$.  
Another advantage of the geometric approach is that it allows to easily identify the less dissipating transformations of $H_t$ by simple inspection of the eigenvalues of the metric. This was illustrated in the qubit case, where the effect of quantum coherence or of different timescales in the Lindbladian could be read out from the metric. 

It should however be kept in mind that the geometric approach provides an approximate solution, which becomes only exact  in the slowly driven limit.  
To quantitatively analyse how good  the approximation is, we now compare it to the exact solution for a quantum dot presented in \cite{esposito2010finite} (the same results can be obtained using \cite{cavina2018variational}) with the one obtained solving the geodesic equation with the same Lindbladian as above and a varying Hamiltonian of the form: $H = \frac{\epsilon(t)}{2}\norbra{\id+\sigma_{z}}$. For simplicity of notation, we will measure the energy in units of $\beta$ and the times in $\tau$. 
In \cite{esposito2010finite} it is shown that optimal protocols will present a quench in the Hamiltonian at the beginning and at the end of the protocol when $T$ is small. The size of the jumps is proportional to the duration of the process, and goes to zero for $T\rightarrow\infty$. This behaviour is illustrated on the left hand side of figure \ref{fig:Comparison}.
Comparing the asymptotic solution for $T\rightarrow\infty$ with the one provided by the geodesic equation we notice that the two actually coincide. As expected, this implies that geodesics become optimal in the limit of big $T$, that is, in the regime in which the quenches at the beginning and at the end of the protocol becomes negligible. This trend is confirmed by the plot on the right hand side of Fig.~\ref{fig:Comparison}, where we plot the total work necessary to excite the quantum dot from $\epsilon(0) = -2 \beta$ to $\epsilon(1) = 2 \beta$ in function of the duration of the protocol. In fact,  the two minimisation protocols give results which are reasonably similar already for times of order $T \approx\tau$, diverging by $\sim 10\% $ when $T = \tau$.

Summarising, both the approach reported here and in \cite{cavina2018variational} allow for minimising dissipation given a Lindbladian master equation, and they appear  to be complementary: \cite{cavina2018variational}  provides exact solutions, whereas the geometric approach provides approximate solutions (which become exact in the slow driving limit)  with a considerably simpler minimisation. The latter approach appears particularly suited for  mesoscopic/many-body systems, where the number of control parameters in the Hamiltonian is small but the Hilbert space dimension is exponentially large.

\end{document}